\documentclass[iop,apj]{emulateapj}

\pdfoutput=1

\usepackage{color}
\usepackage{times}
\usepackage{microtype}
\usepackage{float}

\usepackage{amsmath}
\usepackage{apjfonts}
\usepackage{graphicx}
\usepackage{ifpdf}
\ifpdf
\else
\fi

   \shorttitle{Flux Rope Network in 3D Reconnection}  
   \shortauthors{KAGAN ET AL.}

\begin{document}
\title{A Flux Rope Network and Particle Acceleration \\ in Three Dimensional Relativistic Magnetic
  Reconnection}

\author{Daniel Kagan \altaffilmark{1}, 
Milo\v s Milosavljevi\'c \altaffilmark{1}, and
             Anatoly Spitkovsky \altaffilmark{2}} 

\affil{
 $^1$ Department of Astronomy, University of Texas at Austin,
  Austin, TX 78712 \\
  $^2$ Department of Astrophysical Sciences, Princeton University,
Princeton, NJ 08544}

\begin{abstract}

We investigate magnetic reconnection and particle acceleration in
relativistic pair plasmas with three-dimensional particle-in-cell
(PIC) simulations of a kinetic-scale current sheet 
in a periodic geometry.  We include a guide
field that introduces an inclination between the reconnecting field
lines and 
explore outside-of-the-current sheet magnetizations that
are significantly below those considered by other authors carrying out
similar calculations. Thus our simulations probe the transitional
regime in which the magnetic and plasma pressures are of the same
order of magnitude.  The tearing instability is the dominant mode in the current
sheet for all guide field strengths, while the linear kink mode
is less important even without guide field.  Oblique
modes seem to be suppressed entirely.  In its nonlinear evolution, the 
reconnection layer develops a network of
interconnected and interacting magnetic
flux ropes.   As smaller flux ropes merge into
larger ones, the
reconnection layer evolves toward a
three-dimensional, disordered state in which the resulting flux rope
segments contain magnetic substructure on plasma skin depth scales.
Embedded in the flux ropes, we detect spatially and temporally
intermittent sites of dissipation reflected in peaks in the
parallel electric field.
Magnetic dissipation and particle
acceleration persist until the end of the simulations, with simulations with higher magnetization and lower guide field strength
exhibiting greater and faster energy conversion and particle
energization.  At the end of our largest simulation, the
particle energy spectrum attains a tail extending to high Lorentz
factors that is best modeled with a combination of two additional
thermal components.  We confirm that the primary energization
mechanism is acceleration by the electric field in the X-line
region.  The highest energy
positrons (electrons) are moderately beamed with median angles
$\sim 30^\circ-40^\circ$ relative to (the opposite of) the direction of the initial
current density, but we speculate that reconnection in more highly magnetized
plasmas would give rise to stronger beaming. Lastly, we discuss the
implications of our results for macroscopic reconnection
sites, and which of our results may be expected to hold in systems with higher magnetizations.
 

\end{abstract}
\section{Introduction}

Magnetic reconnection \citep[e.g.,][and references therein]{yamada_magnetic_2010} is of
 interest in diverse areas of astrophysics, yet its mechanics
remains incompletely understood.  A tearing instability is
thought to be necessary to initiate reconnection in reversing
magnetic field configurations, but many astrophysical
plasmas have collisional resistivities
that are insufficient to facilitate its growth.
Interpretations of space plasma
measurements \citep[e.g.,][]{chen_observation_2008,oieroset_direct_2011} and astronomical observations, 
however, suggest that
efficient collisionless reconnection is ubiquitous.
Therefore, collisionless effects, which operate on plasma kinetic scales,
seem to be required to provide the dissipation necessary for effecting
a change of
magnetic topology.  Magnetohydrodynamic (MHD) models for the dynamics of
systems undergoing magnetic reconnection have been available for a long time
\citep[e.g.,][and references therein]{priest_magnetic_2000}.  Unfortunately, these
models do not describe the underlying nature of (possibly multiscale)
plasma organization in the reconnection layer where 
magnetic energy is being dissipated and the assumptions of ideal
MHD do not apply.  Understanding the detailed plasma organization on all length scales,
from the likely relatively small, plasma kinetic scales, to the potentially much larger
scales on which astrophysical dynamical systems ``prepare''
reconnection sites, and where ideal MHD may be valid, is paramount
for completing the theories of a wide variety of astrophysical
phenomena and for
interpreting space plasma measurements and astronomical observations.

In an effort to develop a picture of magnetic reconnection from 
first principles, recent 
particle-in-cell (PIC) simulations
have examined the dynamics of 
reconnection layers that start with a current sheet as thin as the
plasma skin depth.  PIC simulations of reconnection in pair plasmas
have been carried out in two spatial dimensions
\citep{zenitani_generation_2001,zenitani_particle_2007,jaroschek_fast_2004,bessho_collisionless_2005,bessho_fast_2007,bessho_fast_2010,bessho_fast_2012,daughton_collisionless_2007,hesse_dissipation_2007,drake_magnetic_2010,hoshino_stochastic_2012}
and three dimensions
\citep{zenitani_three-dimensional_2005,zenitani_role_2008,yin_three-dimensional_2008,liu_particle_2011,sironi_acceleration_2011}.
Among these, several
\citep{zenitani_three-dimensional_2005,zenitani_role_2008,bessho_fast_2007,hesse_dissipation_2007,hoshino_stochastic_2012}
have investigated the role of departure from the idealized, exactly antiparallel
reconnection by introducing a perpendicular ``guide'' field.  These various
simulations have revealed novel forms of small-scale plasma self-organization that
are interesting in their own right, but that must ultimately be
related to and embedded within the appropriate 
larger astrophysical contexts \citep[e.g.,][]{uzdensky_fast_2010}.
While spacecraft measurements, which can be done in situ, can provide
direct clues how to establish this embedding in space plasmas,
in extrasolar contexts only an indirect relation can be established 
between the reconnection process and the observed emission 
\citep[e.g.,][]{sironi_acceleration_2011,cerutti_beaming_2012}.

Common
features seen in many PIC simulations of magnetic reconnection 
include the formation of chains of
magnetic flux ropes (in three dimensions with a guide field; otherwise, the common terms
``islands'' or ``plasmoids'' may still be more appropriate), the merging of
smaller flux ropes into larger ones, and an energization of the plasma in
the reconnection layer.  In
three dimensional simulations, kink-like and oblique modes, as
well as secondary instabilities, can impart three
dimensional structure to the reconnection layer. 

Typically, the simulations are initialized in the so-called Harris sheet
equilibrium describing a current sheet with
a thickness similar to the plasma skin depth.
The tearing instability first sets in on scales of the initial current
sheet thickness.  Its nonlinear development produces a chain of
skin-depth-scale flux ropes alternating with magnetic X-lines, the
three dimensional generalization of two dimensional X-points.  In
X-lines, violation of flux freezing and magnetic line reconnection
can be 
facilitated by a pressure
tensor anisotropy \citep[][see also, e.g., \citealt{hesse_dissipation_2007},
and references therein]{vasyliunas_theoretical_1975}.
Smaller flux ropes tend to merge with each other to form larger ones; this gives rise to magnetic organization on
increasingly larger spatial
scales.    Three-dimensional
PIC simulations of guide field reconnection in electron-ion plasmas 
exhibit these same features, e.g., \citet{daughton_role_2011} 
found that oblique modes dominated 
over tearing modes when guide field was strong.
Additional effects specific to plasmas with electron-ion mass
disparity have also been identified, but are not relevant for the present work.

Energization of particles in reconnection layers has been investigated in a number of PIC
simulations
\citep{zenitani_generation_2001,zenitani_particle_2007,jaroschek_fast_2004,drake_electron_2006,drake_magnetic_2010,bessho_fast_2007,bessho_fast_2010,bessho_fast_2012,egedal_formation_2009,huang_mechanisms_2010,oka_electron_2010,liu_particle_2011,sironi_acceleration_2011,egedal_large-scale_2012,hoshino_stochastic_2012,cerutti_beaming_2012}.
 Less attention has been given to particle energization in the general
case of reconnection with a guide field in three dimensions
\citep{zenitani_role_2008}.  In two dimensional simulations, particle
acceleration producing a nonthermal energy spectrum, an apparent power-law, is often 
reported. \citet{cerutti_beaming_2012}, however, instead detect a
new ultrarelativistic thermal component energized by the
reconnection.

That the reconnection layers should energize particles
is in agreement with analytical considerations
\citep[e.g.,][]{speiser_particle_1965,larrabee_lepton_2003,giannios_uhecrs_2010,uzdensky_reconnection-powered_2011,cerutti_extreme_2012},
which find that particles in the vicinity of the X-line in the
reconnection layer are accelerated by the nearly-uniform electric
field as they repeatedly cross, and are trapped within the converging
plasma flows.  Other mechanisms focusing on energetic particles that have moved from
the X-line region into the flanking islands have also been suggested
\citep[e.g.,][]{drake_electron_2006,drake_magnetic_2010}.   In three
dimensional simulations, evidence for a nonthermal spectrum is less solid.
It remains poorly 
understood which processes limit the energy to which particles can be accelerated
in fully dynamical, three-dimensional reconnection layers, and how
do the particle energy spectrum, the
degree of accelerated particle beaming, and the temporal evolution of
the accelerated population depend on the parameters of the reconnection layer.

In this work we employ three dimensional PIC simulations 
to investigate the evolution of current sheets in
relativistic pair plasmas undergoing magnetic reconnection.  Our simulations add to the
small but growing family of three dimensional PIC simulation of
relativistic reconnection with a guide field.   With the
intention to complement existing work, we initialize our simulations
slightly differently than it is normally done, not assuming the usual
Harris sheet equilibrium.  Also, 
we explore a parameter regime, involving
magnetic to kinetic pressure ratios of the order of unity, 
that has thus far not received sufficient attention.  We observe an evolution of magnetic
field geometry that constrains the viability
of models
for high-Lundquist-number reconnection layers in which the diffusion
region contains a hierarchy of interacting plasmoids 
\citep[e.g.,][]{shibata_plasmoid-induced_reconnection_2001,uzdensky_fast_2010}. 
The simulations
also allow us to explore the character of particle energization in dynamical, fully
three dimensional reconnection layers.

The paper is organized as follows.  Section \ref{sec:simulations}
describes our methodology and simulation setup, while  Section
\ref{sec:results} presents the results.  Section \ref{sec:discussion}
discusses our findings concerning development of kinetic instabilities
in the current sheet, as well as our findings on particle
energization, in view of the existing work on these topics. Finally,
Section \ref{sec:conclusions} reviews our main conclusions.

\section{Description of Simulations}
\label{sec:simulations}
\subsection{The Initial Configuration}

The spatial domain is rectangular with $0\leq x<L_x$, $0\leq y<L_y$, and $0\leq z<L_z$. The boundary conditions are periodic in all directions.
The initial magnetic field is the same as that in the Harris equilibrium
\begin{eqnarray}
  {\mathbf B}&=&B_0 \left[ \tanh \left(\frac{x-L_x/4}{\lambda_0}\right) - \tanh \left(\frac{x-3L_x/4}{\lambda_0}\right)-1\right]\hat{{\mathbf z}}\nonumber\\& & +\kappa \,B_0\, (-\hat{\mathbf y}) ,
\label{eq:harris_field}
\end{eqnarray} 
where $\lambda_0$ is the half-width of the initial current sheet and $\kappa\geq 0$ is a parameter defining the strength of the uniform guide field perpendicular to the opposing field, which in our simulations is oriented in the $-y$ direction.
The current sheets are located at $x=L_x/4$ and $x=3L_x/4$ and carry antiparallel currents.
The current density profile that satisfies Amp\`ere's law is
\begin{equation}
{\mathbf J}=-\frac{c B_0}{4\pi \lambda_0 }\left[{\rm sech}^2\left( \frac{x-L_x/4}{\lambda_0}\right) - {\rm sech}^2 \left( \frac{x-3L_x/4}{\lambda_0}\right)\right]\hat{{\mathbf y}} .
\label{curprofile}
\end{equation}

To ensure that Amp\`ere's law is satisfied, if we set the particle
density to be uniform, the particles have a spatially-dependent drift velocity $\boldsymbol{\beta}_i=-\boldsymbol{\beta}_e=\boldsymbol{\beta}$ in the $-y$-direction. The current density ${\mathbf J}$ is related to the velocity $\boldsymbol{\beta}$ by the relation
\begin{equation}
{\mathbf J}=n_0 e c(\boldsymbol{\beta}_i-\boldsymbol{\beta}_e)=2 n_0 e c \boldsymbol{\beta} ,
\label{curdrift}
\end{equation}
where $e$ is magnitude of the unit charge carried by electron and ion macroparticles.
This results in the drift velocity profile
\begin{equation}
\label{eq:beta_profile}
\boldsymbol{\beta}=-\beta_0 \left[{\rm sech}^2\left( \frac{x-L_x/4}{\lambda_0}\right) - {\rm sech}^2 \left( \frac{x-3L_x/4}{\lambda_0}\right)\right]\hat{{\mathbf y}} ,
\end{equation}
where $\beta_0 \equiv B_0/(8\pi n_0 e \lambda_0)$. 

 In this work, we initialize the
simulation outside of pressure equilibrium with a uniform initial
density $n_0$ and spatially varying drift velocity $\boldsymbol{\beta}$.  We can still attempt to
relate the parameters of our simulation to those of preceding
investigations. In setting up initial conditions for a plasma with the magnetic field
given in Equation (\ref{eq:harris_field}), there are multiple
ways in which pressure equilibrium can be satisfied
depending on the spatial variation of the plasma density $n_{e^+}+n_{e^-}$,
temperature $T$, charge drift velocity $\boldsymbol{\beta}$, reversing
field strength $B_0$, and
guide field strength $B_y$.  It
is common to assume that the initial temperature is uniform and equal
to $T_0$, and only the density varies across the current sheet.  In
practice, Harris sheets are often set up with a strong excess density
in the current sheet for pressure balance.
It is common to split the particle population into two components, 
one a uniform background with density $n_{\rm b}$,
and another spatially varying with maximum density $n_0$ at the center
of the current sheet.  The latter component ensures pressure
equilibrium and carries the
current in the reconnection layer.  In Section
\ref{sec:equilibrium} below, we discuss how the initial configuration
adjusts to approximate pressure
equilibrium.

One can define the
magnetic to kinetic pressure ratio via
\begin{equation}
\sigma\equiv\frac{P_{\rm mag}}{P_{\rm kin}}=\frac{B^2}{8 \pi (n_{e^+}+n_{e^-}) T} ,
\end{equation}
where here and henceforth
we express the temperature in energy units.
Then in the isothermal Harris sheet pressure equilibria used in preceding
investigations, 
the ratio simply
equals the density contrast in the current sheet, $\sigma=n_0 /n_{\rm
  b}$.  In our simulation, $\sigma$ is uniform outside the current
sheet and equals
\begin{equation}
\label{eq:sigma_uniform}
\sigma=\frac{B_0^2}{16 \pi n_0 T_0} . 
\end{equation} 
Note that $\sigma$ is defined not taking into account the magnetic
pressure of the guide field.

We do not introduce any initial perturbation to the initial field
geometry described here.  The structure that develops is thus seeded
by numerical fluctuations.

\subsection{Parameters}

We initialize the simulation at temperature $T_0=m_e c^2$. 
All the particles are drawn from the relativistic Maxwellian
distribution, implying that the average kinetic energy
of the particles is $\sim2.37\,m_e c^2$.
We are interested in the dependence of reconnection mechanics and
evolution of particle energy distribution on the dimensionless ratio
of the magnetic pressure to the particle pressure $\sigma$, and the
guide field amplitude parameter $\kappa$.   A nonvanishing $\kappa$
indicates that the magnetic field is twisted in the current sheet and
the plasma in the sheet center is magnetized.  We run a grid of nine simulations,
with $0.25\leq\sigma\leq 2$ and $0\leq \kappa\leq 1$, and carry out a
detailed study of the run with $\sigma=2$ and $\kappa=0.25$. The
chosen values of $\sigma$ are low because we are interested in finding
the lower limit of magnetization at which reconnection can produce
significant particle energization.

\begin{deluxetable*}{lccccccccccc}
 \setlength{\tabcolsep}{0.00in}
 \tablewidth{6.5in}
  \tablecolumns{12}
  \tablecaption{Simulation Parameters and Results\label{tab:simulations}}
\tablehead{\colhead{Run} & \colhead{$L_x$\ \tablenotemark{a}} &
  \colhead{$L_{y}$\ \tablenotemark{a}}  & \colhead{$L_{z}$\
    \tablenotemark{a}} &\colhead{$\lambda_0$\ \tablenotemark{a}} &
  \colhead{$\sigma$} & \colhead{$\kappa$} & \colhead{$f_{\rm c}$\
    \tablenotemark{b}} & \colhead{$|\Delta {\mathcal E}_B|/{\mathcal E}_B$\,(\%)\
    \tablenotemark{c}} & \colhead{$K_{\rm ener}/K$\,(\%)\ \tablenotemark{d}}
  & \colhead{$K_{\rm ener}/|\Delta {\mathcal E}_B|$\,(\%)\ \tablenotemark{e}} & \colhead{$\gamma_{\rm max}$\ \tablenotemark{f}}}
\startdata
{\tt S1K0}  &64&40&40&2&1& 0 &1.5&72&$3.1$& 11 & 48.4\\
{\tt S1K025}  &64&40&40&2&1& $0.25$&1.0&28&$3.1$& 23 & 34.3 \\
{\tt S1K1}  &64&40&40&2&1& $1$&1.0&4.3&$1.5$&  32 & 34.2\\
{\tt S2K0}  &64&60&60&3&2& 0&2.3&55&$10$&  30 & 52.6 \\
{\tt S2K025}  &64&60&60&3&2& $0.25$ &2.3&44&$9.3$& 29  & 54.4\\
{\tt S2K025L} &128&120&120&3&2& $0.25$&2.3&27 &$9.1$& 38 & 58.9\\
{\tt S2K025P}\tablenotemark{g} &64&60&60&3&2& $0.25$ &2.3&40&$12$&  18 & 115\\
{\tt S2K05} &64&60&60&3&2& $0.5$ &2.3&25&$6.4$& 27   & 44.9\\	
{\tt S2K1}  &64&60&60&3&2& $1$&2.3&6.7&$3.2$&  28 & 42.8
\enddata
\tablenotetext{a}{The length scales $L_x$, $L_y$, $L_z$, and
  $\lambda_0$ are given in units of the plasma skin depth $\lambda_{\rm p}$.}
\tablenotetext{b}{$f_{\rm c} $ is the ratio of the original current sheet width to the current sheet width after the readjustment phase.}
\tablenotetext{c}{The ratio $|\Delta {\mathcal E}_B|/{\mathcal E}_B$ is the fraction of
  magnetic energy converted into kinetic energy during the
  simulation.}
\tablenotetext{d}{$K_{\rm ener}/K$ is the particle kinetic energy
  fraction in energized particles (see Section \ref{sec:energization_efficiencies}).}
\tablenotetext{e}{$K_{\rm ener}/|\Delta {\mathcal E}_B|$ is the ratio of the
  particle kinetic energy in energized particles to converted magnetic
  energy (see Section \ref{sec:energization_efficiencies}).}
\tablenotetext{f}{$\gamma_{\rm max}$ is the Lorentz factor of the
  highest energy particle in the simulation.}
  \tablenotetext{g}{In this run an accelerated population with equal numbers having $\gamma=40,60,$ and 80 is added to the thermal population at the beginning of the simulation; only the value of $\gamma_{\rm max}$ changes significantly as a result.}
\end{deluxetable*}

Because of the low growth rate, the simulations with $\sigma=0.25$
did not develop the tearing instability over the time period of the
simulations and thus did not undergo
reconnection, therefore, we do not show the results of these simulations
in what follows.  The parameters of the simulations that
did undergo reconnection are shown
in Table \ref{tab:simulations}.

\subsection{Simulation Method and Resolution Requirements}

We use the relativistic particle-in-cell (PIC) plasma code {\tt TRISTAN-MP} \citep{spitkovsky_structure_2008} to simulate the evolution of a
reconnection configuration in a pair plasma. In a PIC simulation, the number of macroparticles of each species located in each grid cell must be large enough to resolve variations in the current density and limit high-frequency particle noise. {\tt TRISTAN-MP} uses a current filtering algorithm to reduce high-frequency particle noise, substantially reducing the required number of macroparticles per cell per species. We initialize our simulations with 4 macroparticles per cell per species, which is fewer than found in other recent PIC simulations \citep{daughton_role_2011,liu_particle_2011}. To verify that this low particle density does not lead to cell evacuation, we calculate the total particle density including both species in each cell in Run {\tt S1K025} during the flux rope merging phase discussed in Section \ref{sec:results} below. We find that the total particle density in each cell calculated with a cubic cloud-in-cell kernel of size equal to the grid spacing is always larger than 1, with a total density larger than $5$ in over $ 96 \%$ of cells. Cells with low particle density occur in regions characterized by weak spatial field gradients that should not be sensitive to particle noise. The cells with strong field gradients that would be sensitive to particle noise invariably have a higher than average particle density.

To ensure that the particle density is indeed sufficient to resolve the physics of reconnection, we have carried out two dimensional convergence tests, as well as a longer three dimensional test run with 80 particles per cell per species.  In our simulations, the tearing instability leading to reconnection is seeded by numerical fluctuations that vary from simulation to simulation, so we do not expect all observables to be the same in each run at a given absolute time measured from the beginning of the simulation. We therefore compare measurable quantities in each simulation at similar points in their evolution; typically, we calculate these quantities at the end of the flux rope merging phase, which is discussed in Section \ref{sec:results} below. The coarse time discretization employed in recording simulation output makes the determination of these evolution points imprecise. This results in inaccuracies in the estimation of observables that vary quickly with time; such quantities will therefore vary with resolution even if the resolution is sufficient to capture the physics of reconnection. 

Our two dimensional convergence tests use initial conditions matching those in the three dimensional simulations with $\sigma=2$ and $\kappa=0$, but with a larger number, up to 16, of macroparticles per species per cell. We find that the only significant discrepancy, up to $40\%$, is seen in the peak linear growth rate of the kinetic tearing instability; this quantity is very sensitive to the coarse time discretization because time differencing is required for growth rate calculation. To test that we have sufficient macroparticle density to account for 3D effects in our simulations, we also carry out a three dimensional simulation with 80 macroparticles per cell per species, and the same parameters as in Run {\tt S2K025} with $\sigma=2$ and $\kappa=0.25$. We again find that the measurable quantities do not deviate greatly from those found in Run {\tt S2K025}, with the maximum deviation of $30\%$ again seen in the growth rate of the kinetic tearing instability. We also find that the evolution of the total magnetic energy and the particle energy spectrum in this simulation with higher macroparticle density does not differ greatly from that found in Run {\tt S2K025} in Section \ref{sec:energization}. The main difference is that in the high-density simulation, the flux rope merging phase identified in Section \ref{sec:results} begins slightly later; this may be understood as the result of slightly lower particle noise in the high density simulations, which results in later growth of instability. 

To ensure that our simulations resolve the plasma skin depth, we set
$\Delta x=\lambda_{\rm p}/8$, where the skin depth is given by 
\begin{equation}
\lambda_{\rm p}=\sqrt{\frac{\left\langle \gamma\right\rangle m_e c^2 }{8 \pi n_0 e^2 }} .
\label{plasma}
\end{equation}
Here, $\left\langle \gamma\right\rangle$ is the average Lorentz factor
of particles in the simulation and the additional factor of 2 in the
denominator reflects the fact that the electrons and positrons oscillate
together. 
To determine whether this value for $\lambda_{\rm p }$ is sufficient to resolve the physics of reconnection, we have carried out further two dimensional convergence test simulations with up to 3 times larger
number of grid cells per skin depth than in the three dimensional
simulations, and again compared results at the end of the flux rope merging phase.  Here, the most resolution-dependent quantity was the
maximum value of the reconnected magnetic field $B_x$, which varied by
$25\%$ with resolution. This is a quantity that varies quickly with time near the end of the flux rope merging phase, so the apparent variation with resolution is again likely a result of coarse time discretization.

We focus on the regime in which the particle drift velocity in Equation
(\ref{eq:beta_profile}) giving rise to the
current density in
Equation (\ref{curprofile}) is nonrelativistic. This places a constraint on the initial current sheet width $\lambda_0$.  Combining Equations
(\ref{curprofile}) and (\ref{curdrift}) with the definition of
$\sigma$ in Equation (\ref{eq:sigma_uniform}), it can be shown that
\begin{equation}
\frac{\lambda_0}{\lambda_{\rm p}}=\frac{\sqrt{2 \sigma (\Gamma-1)}}{\beta_0} ,
\end{equation}
where $\Gamma-1\equiv T_0/(\left\langle \gamma\right\rangle m_e c^2)$
is the ratio of the particle pressure to the particle energy density
(which includes the particle rest energy); in our simulations,
$\Gamma-1 \approx 0.297$.\footnote{The index $\Gamma$ is related to the
temperature via
\begin{equation}
\frac{1}{\Gamma-1} = \frac{1}{\theta} \left[\frac{K_3(\theta^{-1})}{K_2(\theta^{-1})}-\theta\right] ,
\end{equation} 
where $\theta\equiv T/(m_e c^2)$ and $K$ is the modified Bessel
function of the second kind.}
With this estimate of the relation between $\lambda_0$ and $\beta_0$, in simulations with $\sigma\leq 1$ we choose $\lambda_0/\lambda_{\rm
  p}=2$, while in simulations with $\sigma=2$ we choose
$\lambda_0/\lambda_{\rm p}=3$.  This implies that
$\beta_0\approx 0.35$, 
ensuring that the drift velocities are not relativistic.

\subsection{Readjustment to Equilibrium}
\label{sec:equilibrium}

The uniform initial density and temperature and nonuniform magnetic
field in the initial configuration defined by Equations
(\ref{eq:harris_field}), (\ref{curdrift}), and (\ref{eq:beta_profile})
are not in force balance, but they achieve approximate force balance
following a brief readjustment.
 The
initial adjustment to equilibrium results in a reduction of the current
sheet thickness and an accompanying compression of the plasma by a factor
\begin{equation}
f_{\rm c} \equiv \frac{\lambda_0}{\lambda} > 1,
\end{equation}
where henceforth, $\lambda$ denotes the current sheet half-thickness after
the readjustment.  We further define 
\begin{equation}
\tau_{c0} \equiv \frac{\lambda_0}{c} , \ \ \ 
\tau_c \equiv \frac{\lambda}{c} 
\end{equation}
to denote the light crossing times of the initial and the compressed current sheet.

The plasma takes $\approx10\,\tau_{c0}$ to adjust to a state close to pressure
equilibrium.
The magnitude of the compression is shown in
Table \ref{tab:simulations}. With readjustment, both the density of the particles and
the strength of the guide field increase in the center of the
current sheet in such a way as to balance the large pressure gradient
that the reversing field exhibits from the center to the surface of the current
sheet.
Longer-term transients are present in the form of acoustic
oscillations. 
These oscillations could have the effect of
enhancing the tearing instability, which relies on a velocity inflow
to the current sheet. The initial readjustment does not give rise to
any significant magnetic dissipation or particle energization.

\subsection{Unstable Modes and the Size of the Simulation}
\label{sec:unstable_modes}

For a simulation to capture the physics of reconnection, its size must
be sufficient to include the fastest growing modes of the important
instabilities of relativistic Harris current sheets in pair plasmas.
The known instabilities include the kinetic relativistic tearing
instability (KTI; hereafter the tearing mode), the drift-kink
instability (DKI; hereafter the kink mode), and the oblique mode
which is similar to the tearing mode. To estimate
their wavelengths, we follow \citet{zenitani_particle_2007}. The
growth rate $\omega_{\rm KTI}$ of the tearing mode with wave number
$k_z$ in a Harris current sheet of half-thickness $\lambda$ and non-relativistic drift velocity $\beta$ is given by
 \begin{equation}
        \omega_{\rm KTI} = b(T)\,\beta^{3/2} \,k_z\lambda\,[1-(k_z \lambda)^2]\,\frac{1}{\tau_c},
        \label{eq:tearing}
\end{equation}
where is a dimensionless function of the plasma
temperature that in the limits of a cold and a relativistically hot
plasma equals
\begin{equation}
b(T) = \begin{cases}
\frac{1}{\sqrt{\pi}}  \left( \frac{2T}{\gamma_\beta
    m_e c^2}\right)^{-1/2}  & ,\ \   T \ll m_e c^2 , \\
\frac{2\sqrt{2}}{\pi}  & ,\ \  T \gg m_e c^2 ,
\end{cases}
\end{equation}
and $\gamma_\beta\equiv (1-\beta^2)^{-1/2}$. 
The resulting maximum growth rate occurs for $k_z
\lambda=1/\sqrt{3}$, corresponding to a wavelength of
$\sim 10.8\, \lambda$.  If we use the form of $b(T)$ appropriate at
ultrarelativistic temperatures, the growth rate for this mode is 
\begin{equation}
\label{eq:KTI_max}
\omega_{{\rm KTI},{\rm max}} = 0.35\, \frac{\beta^{3/2}}{\tau_c} .
\end{equation}

The situation is more complicated for the kink mode, because as
\citet{zenitani_particle_2007} find, the analytical maximum value of
the growth rate $\omega_{\rm DKI}$ occurs for $k_y\lambda>1$, a
wavenumber for which kinetic effects become important in a thin
current sheet. Simulation results in their Figure 20 indicate that the
maximum growth rate occurs at $k_y\lambda\approx 0.7$ corresponding to a
wavelength of $\approx 9\,\lambda$.

In addition to these two dimensional modes, it is also necessary to resolve oblique
modes with $k_y, k_z\neq0$ that combine tearing and kink components; these
were identified in three dimensional simulations by
\citet{zenitani_three-dimensional_2005,zenitani_role_2008} and
\citet{daughton_role_2011}.  The typical fastest-growing oblique mode
in both of these simulations had $k \lambda\sim 0.2$; this corresponds
to a wavelength in both directions of $\sim 30\, \lambda$.

To resolve all three types of modes, in all
simulations except for the larger size run {\tt S2K025L}, we set
$L_y=L_z=20\,\lambda_0$, large enough to contain at least one wavelength
of the tearing mode and two wavelengths of the kink mode. Because the initial
adjustment leads to a significant narrowing of the current sheet,
several wavelengths of the fastest-growing tearing and drift-kink
modes of the narrower current sheet are included in the simulations,
and at least one wavelength of the oblique modes should also be
resolved. The resulting overall length scales are $L_x =
64\,\lambda_{\rm p}$ and $L_y=L_z=40\,\lambda_{\rm p}$ or
$L_y=L_z=60\,\lambda_{\rm p}$, depending on the simulation, as is
shown in Table \ref{tab:simulations}. To ensure that little
interaction takes place between the two current sheets in the periodic
box, we set $L_x = 64\, \lambda_{\rm p}$.  The current sheets in our larger size simulation
{\tt S2K025L} are as thick as in the other ones, but their separation and
the dimensions of the box are twice as large, i.e., $L_x=128
\,\lambda_{\rm p}$ and $L_y=L_z=120\,\lambda_{\rm p}$.  
This simulation was carried out on a $1024\times960\times960$ grid and contained a total of $\approx 7.5\times10^9$
particles. The smaller simulations with $\sigma=1$ were carried out on
a $512\times320\times320$ grid and contained $4.2\times10^8$ particles,
whereas the simulations with $\sigma=2$ were carried out on a $512\times480\times480$ grid with
$9.4\times10^8$ particles.

\subsection{Duration of the Simulations}

In our simulations the growth times of the tearing
and kink modes are on the same order as and smaller than the Alfv\'en crossing time of the box.
To capture the physics of reconnection in its nonlinear
regime and on
length scales similar to the box size, the duration of the simulation
must be larger than these time scales.  The 
 relativistic Alfv\'en velocity is given by 
\begin{equation}
	\label{eq:alfven}
	v_{\rm A}=\frac{c}{\sqrt{1+\Gamma/[2 \sigma (\Gamma-1)  ] }} .
\end{equation}
The Alfv\'en crossing time of the box $\tau_{{\rm
    A},z}=L_z/v_{\rm A}$ is $\sim28\,\tau_{c0}$ in the runs with
$L_z=60\,\lambda_{\rm p}$.
We run all of our simulations for at least $8000$ time steps, which
amounts to $150\,\tau_{c0}$ in the large simulation {\tt S2K025L}.

\section{Results}
\label{sec:results}

The evolution of the reconnection layer in all simulations exhibits
the same common that have been observed in previous PIC simulations of
magnetic reconnection in pair plasmas. The tearing instability grows
and produces a chain of alternating X-lines and flux ropes.  Here, we
adopt ``flux ropes'' to denote a magnetic structure sometimes referred
to as ``plasmoid'' (or ``island'' in two dimensional treatments) 
in which plasma is pinched by a helical magnetic
field which can have a braided structure.  The flux
ropes can have a finite lengths, with the field lines opening up and
extending arbitrarily far from the rope axis.   The flux ropes can also split
into sub-ropes, which opens the possibility of an organization of the
ropes in a flux rope network.  In
all simulations except for {\tt S1K0} where the kink instability
disrupts the current sheet, the flux ropes merge in quasi-hierarchical
fashion until either only one flux rope is left in each current sheet
or, in the large simulation {\tt S2K025L}, the flux ropes have been
disrupted by a transition to a disordered, three-dimensional state.
The period of flux rope merging is accompanied by fast
magnetic-to-kinetic energy conversion. This conversion produces a tail
of energized particles.  

In simulations without guide field, the kink instability also grows,
resulting in some corrugation of the current sheet.  For $\sigma=2$ in
simulation {\tt S2K0}, the kink instability does not disrupt the
quasi-hierarchical merging process. For $\sigma=1$ in simulation {\tt
  S1K0}, however, the kinking disrupts the merging and brings the
two current sheets into contact where they can interact; this is
accompanied by rapid conversion of magnetic to thermal energy but the
resulting spectrum does exhibit signatures of a secondary energized
component.   In agreement with \citet{zenitani_role_2008}, we find
that that the presence of a guide field suppresses the kink
instability.  However we find that in the large simulation {\tt
  S2K025L}, a transition to three-dimensional
evolution still occurs, likely due to a lack of large-scale phase coherence in the
tearing instability.  

We proceed to discuss our results in detail.  In Section
\ref{sec:network} we discuss the evolution of the global current sheet and
the formation of what will turn out to be a network of interconnected
flux ropes in our largest simulation.
In Section \ref{sec:dissipation_flux_ropes} we identify sites of
magnetic reconnection within the network.
  In Section \ref{sec:mode_analysis} we
carry out a Fourier decomposition of perturbations in the largest simulation and
discuss their growth rates. In Section \ref{sec:timescales} we analyze
the time scales associated with the evolution of the flux rope
network. In Section \ref{sec:reconnection_rate}
we measure the overall reconnection rate, while in
Section \ref{sec:ohms_law} we analyze
components of the nonideal electric field as reflected in the
generalized Ohm's law. Finally, in Section \ref{sec:energization}
 we discuss the overall rate of magnetic-to-thermal
energy conversion as well as the
efficiencies, mechanisms, and properties of particle energization in
the simulations.

\subsection{Formation and Evolution of the Flux Rope Network}
\label{sec:network}

\begin{figure*}
\begin{center}
\includegraphics[width =0.9\textwidth]{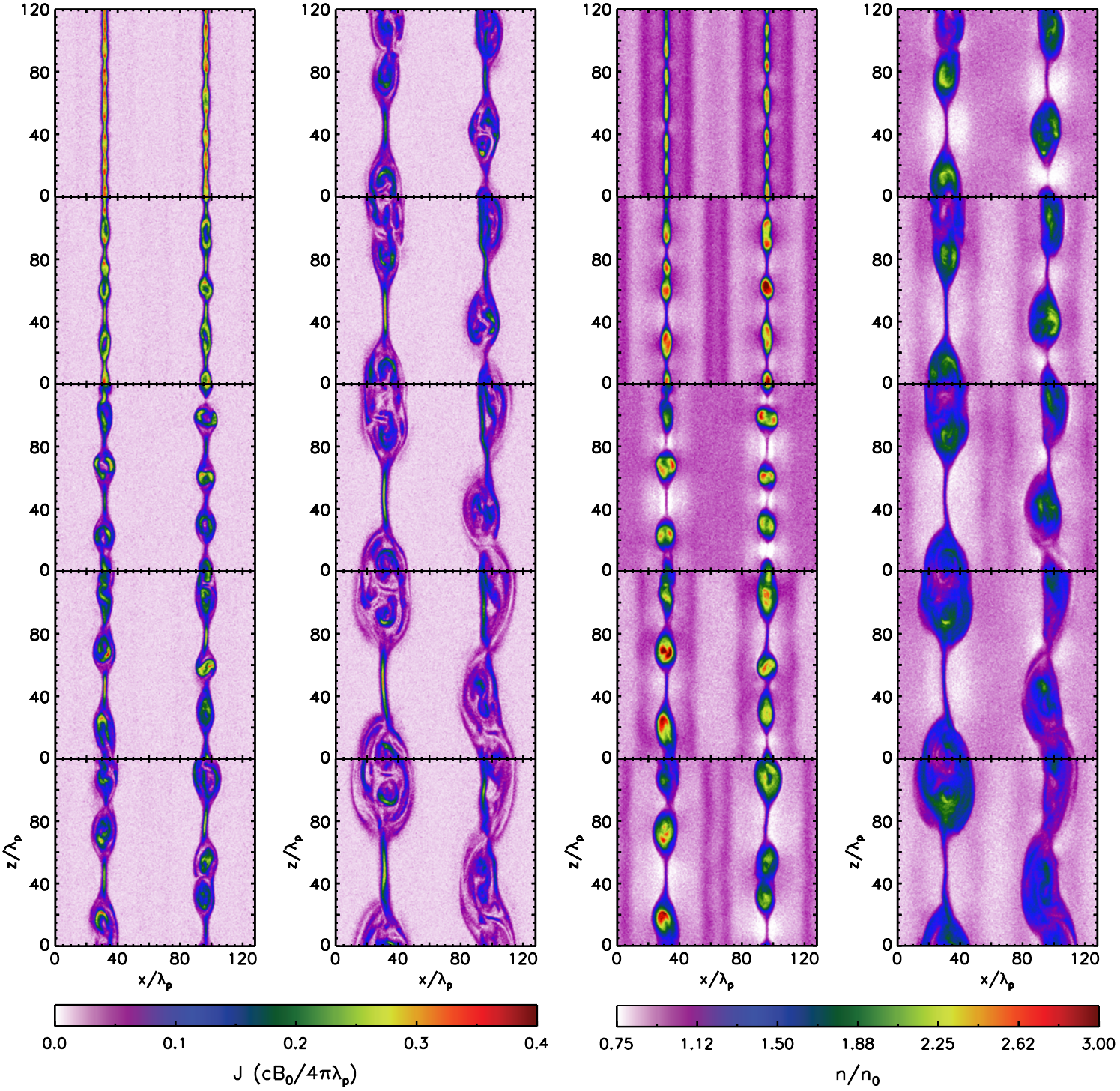}
\end{center}
\caption{The total current density $J\equiv |{\mathbf J}|$ (left two columns) and total particle
  density $n$ (right two columns) in a slice $y=5\,\lambda_{\rm p}$. T
  In each pair of columns, the first column shows times $t=(37.5,\,47,\,56,\,65.5,\,75)\,\tau_{c0}$ and the
  second column shows
  $t=(84.5,\,94,\,103,\,112.5,\,122)\,\tau_{c0}$. The particle density $n$, which is not used directly in the PIC code, is calculated from particle positions using a uniform kernel of half-width 2 grid lengths.   The same kernel is used to smooth the current density $J$.\label{fig:slices}}
\end{figure*}

Figure \ref{fig:slices} shows a time sequence of slices at $y=5\,\lambda_{\rm p}$ of the
total current density $J\equiv |{\mathbf J}|$ and the plasma number 
density $n$ in simulation {\tt S2K025L}.  Outside the two evolving
current sheets,
plane-parallel acoustic waves resulting from the initial pressure
imbalance can be seen in the plasma density. The waves traverse the computational box several times in the
course of the simulation.  The early evolution of each current sheet,
if observed at a single value of $y$, 
is that correctly described by the familiar flux rope (or island, plasmoid)
merging paradigm. Larger flux ropes produced by the merging of smaller
ones contain substructure reflected in multiple, curved, embedded
current sheets, each 
with a half-thickness $\sim \lambda_{\rm p}$.  Very similar
substructure also appears in the recent two-dimensional PIC
simulations of an electron-ion plasma by
\citet{markidis_collisionless_2012}.

\begin{figure*}
\begin{center}
\includegraphics[width = \textwidth]{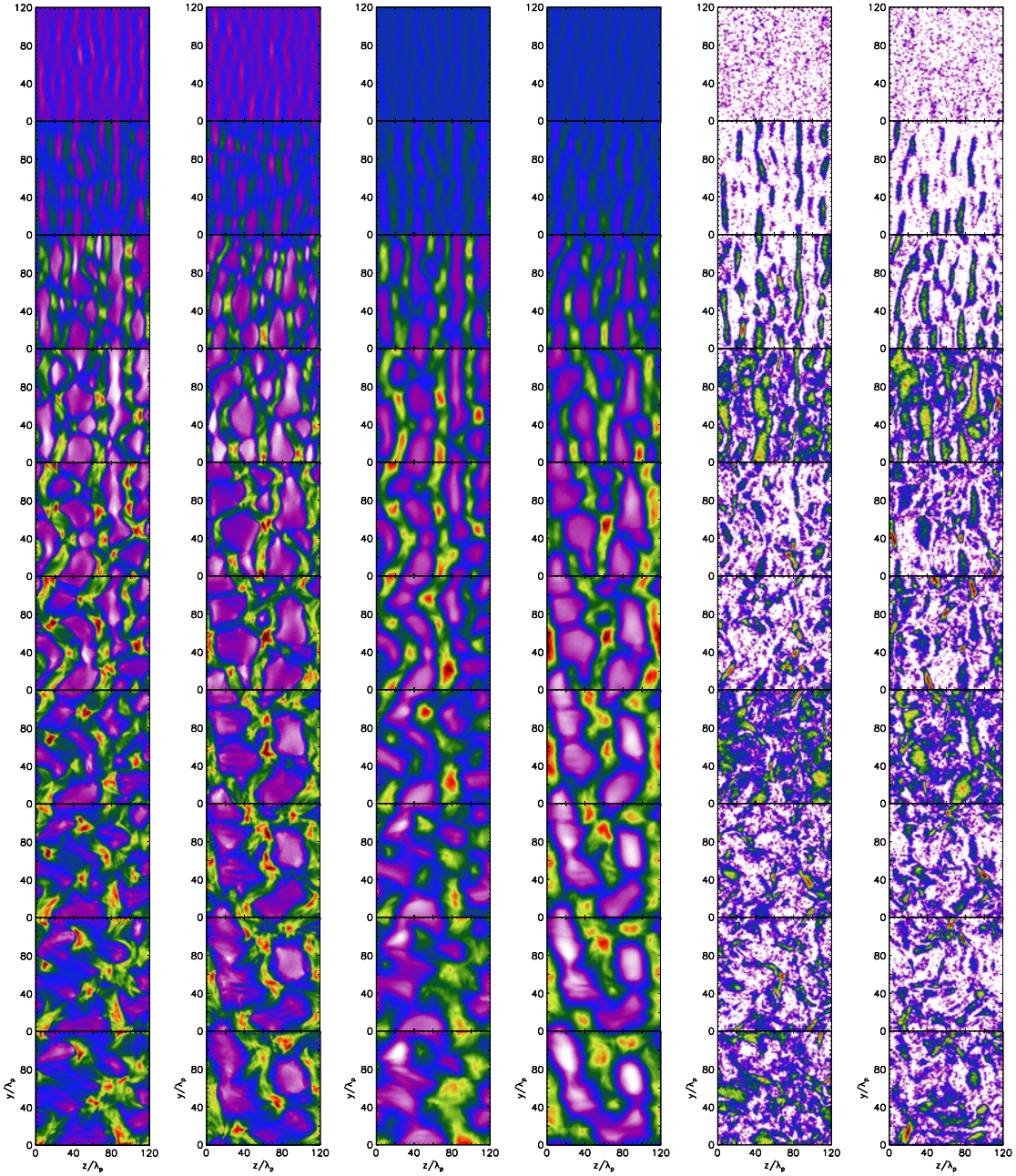}
\end{center}
\caption{The total current density $j\equiv |{\mathbf j}|$ (left two columns), total particle
  density $n$ (middle two columns), and the parallel electric field
  magnitude $E_\parallel\equiv |{\mathbf E}\cdot{\mathbf B}|/B$
  projected onto the $yz$ plane.  In each pair of columns, the left
  column shows
  a projection of the sections $0<x<64\,\lambda_{\rm p}$ containing one
  current sheet, and the right shows a projection of the section 
  $64\,\lambda_{\rm p}< x< 128\,\lambda_{\rm p}$ containing the other
  current sheet.
  Vertically from top to bottom, the panels show times
  $t=(28,\,37.5,\,47,\,56,\,65.5,\,75,\,84.5,\,94,\,103,\,112.5)\,\tau_{c0}$. The
  color scale increasing from light purple to dark red is
  linear except for the parallel electric field, where the scale is
  logarithmic.  The color scale coverage of each projected quantity
  was reduced from the full variation of that quantity for enhanced
  visual contrast.  In particular, clipping of the color scale for the parallel
  electric field excludes the range in which the parallel electric field
  projection is dominated by small-scale electrostatic fluctuations in
  the plasma.   We boxcar smoothed ${\mathbf E}$ and ${\mathbf B}$ on
  scales $\lesssim \lambda_{\rm p}$ prior to computing the parallel
  electric field, but the fluctuations still obscure the variation
  of parallel field in the current sheet at the earliest time shown, $t=28\,\tau_{c0}$.
\label{fig:projections}}
\end{figure*}

To examine three dimensional aspects of the current sheet evolution,
in Figure \ref{fig:projections} we show projections into the $yz$ plane of
the total current density $J$, plasma number density $n$,
and the parallel electric field
  magnitude $E_\parallel$
in each of the two current sheets in simulation {\tt S2K025L}. The
flux ropes that initially develop from
the linear tearing instability are approximately parallel to the $y$
axis, which is the direction of the
initial current flow.  However, each flux rope exhibits one or more
discontinuities where the $z$ coordinate of the rope suddenly
changes.  We interpret these discontinuities as arising from a
lack of tearing mode phase coherence on scales
larger than those in causal contact during the initial
development of the instability.  The lack of phase coherence should
occur on scales
\begin{equation}
\label{eq:coherence_length}
\Delta y, \, \Delta z \gg \lambda_{\rm coh} \sim \frac{c}{\omega}
\end{equation} 
on which locations on the
current sheet are out of causal contact during the first $e$-folding
of the instability; here, $\omega$ is the linear
growth rate of a mode growing from numerical noise.  This sets a hard upper
limit on the scale on which an instability can grow coherently; other constraints may
limit this scale further.  

The initial departure from perfect
translational invariance in the $y$ direction implies that the nonlinear flux rope merging will
itself not occur coherently. In any three adjacent flux ropes, the
middle rope can merge with one of the flanking ropes at one $y$, and
with the other rope at another $y$; at still other values of $y$, the
three ropes may remain separate for a time being.  This creates
lateral linkage between the
flux ropes; already at time $t=37.5\,\tau_{c0}$, all the flux ropes are
mutually linked inside their current sheet.  The linkage implies each
flux rope experiences a strongly $y$-dependent magnetic tension
force.  This magnetic tension from neighboring flux ropes leads to
rope tilting and kinking that is very distinct in origin from that
arising from the well-known linear oblique and kink instabilities of a
current sheet. 

Pairs of large flux ropes emerging from
two generations of flux rope merging, but still oriented largely
parallel to
the $y$ axis, are often connected by minor flux ropes that are highly
tilted toward the $z$ axis. \citet{daughton_role_2011} have discovered
similar structures in a large three-dimensional simulation of
electron-ion reconnection with a strong guide field. It is clear in
Figure \ref{fig:projections} that by $t\sim 75\,\tau_{c0}$, all flux rope
segments have substantial tilts and any semblance of translational
invariance in the direction of the original current flow is lost.  Vertices
of the flux rope network contain highly localized, intense current
sheets and filaments. By $t\sim 122\,\tau_{c0}$, the flux ropes are
largely completely disrupted and the current sheet contains a
disordered network of knot- and sheet-like structures.  

The force ${\mathbf J}\times{\mathbf
  B}$ peaks at the primary X-line current sheets as well as flux rope
perimeters, but is on average much smaller in flux rope interiors. This suggests
that the flux rope interiors are organized in a state of force free
quasi-equilibrium constrained by a nonvanishing magnetic helicity (i.e., twisting,
braiding of the field lines).

\subsection{Dissipation and Reconnection in the Flux Rope Network}
\label{sec:dissipation_flux_ropes}

To gain further insight into the structure of the flux rope network,
we examine the spatial variation of ${\mathbf E}_\parallel$, the component of the electric
field parallel to the magnetic field.  The parallel field vanishes for the purely inductive electric field of
ideal MHD and thus it can be obtained purely from the non-ideal,
reconnection electric field ${\mathbf
  R}$ defined as the difference between the actual and the induction electric field,
\begin{equation}
\label{eq:reconnection_field}
{\mathbf R}\equiv {\mathbf E} +\frac{1}{c}( {\mathbf v}\times {\mathbf B}) .
\end{equation}
 In high magnetic Reynolds number plasmas, 
the parallel field is present only in magnetic diffusion regions, and
in particular, in locations where a change of
magnetic connectivity is under way \citep[e.g.,][]{schindler_general_1988}.\footnote{On very
  small scales in the simulation, ${\mathbf E}_\parallel$ exhibits noise in the entire domain due to small scale
electrostatic fluctuations; in our analysis, we filter these fluctuations on scales
$\lesssim 2\,\lambda_{\rm p}$.}  The parallel electric field
may also be instrumental in maintaining the pressure
anisotropy, 
that is responsible for the breaking of flux freezing in the
reconnection layer \citep[see, e.g.,][]{hesse_diffusion_2011}.

\begin{figure}
\begin{center}
\includegraphics[width = 0.45\textwidth]{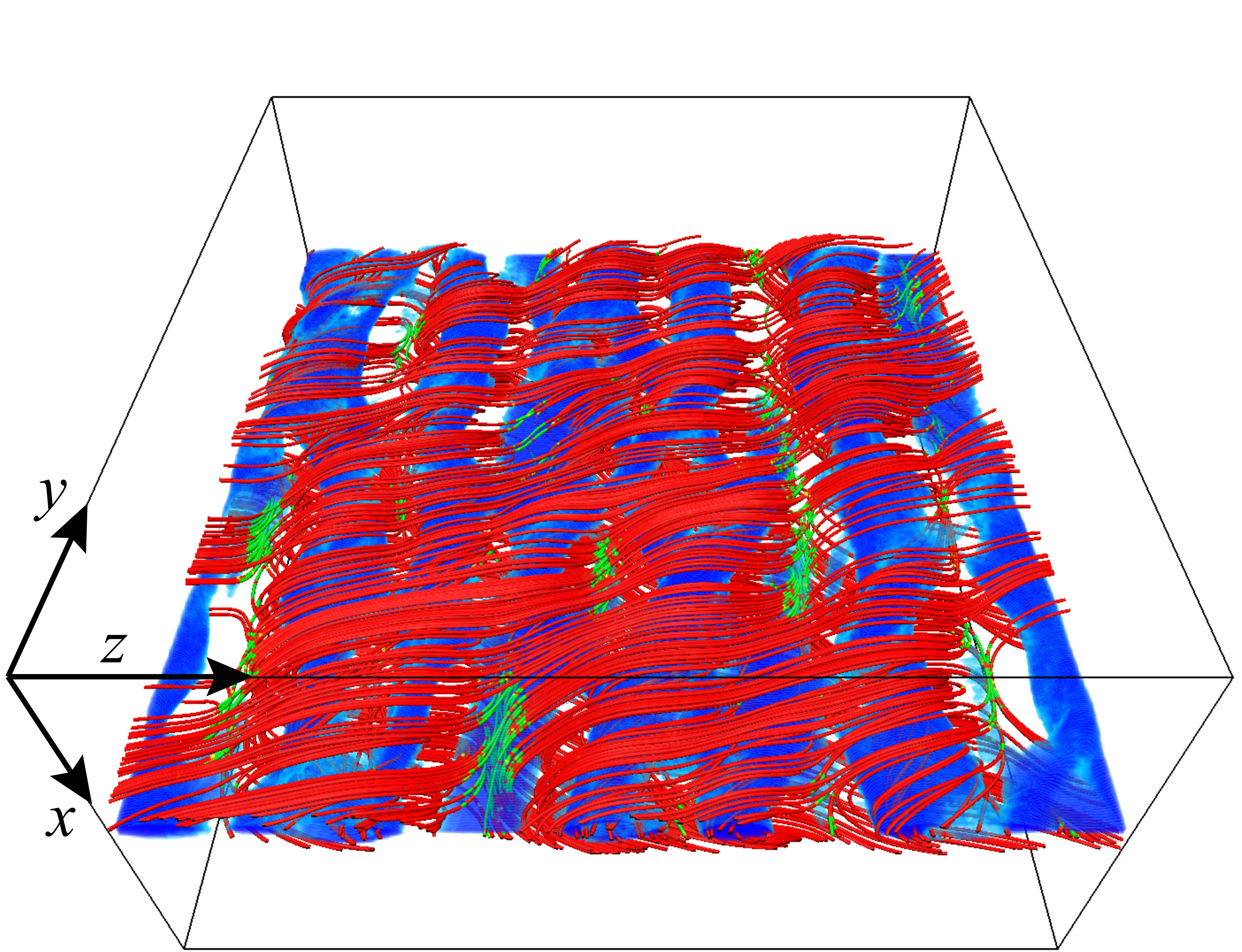}
\includegraphics[width = 0.45\textwidth]{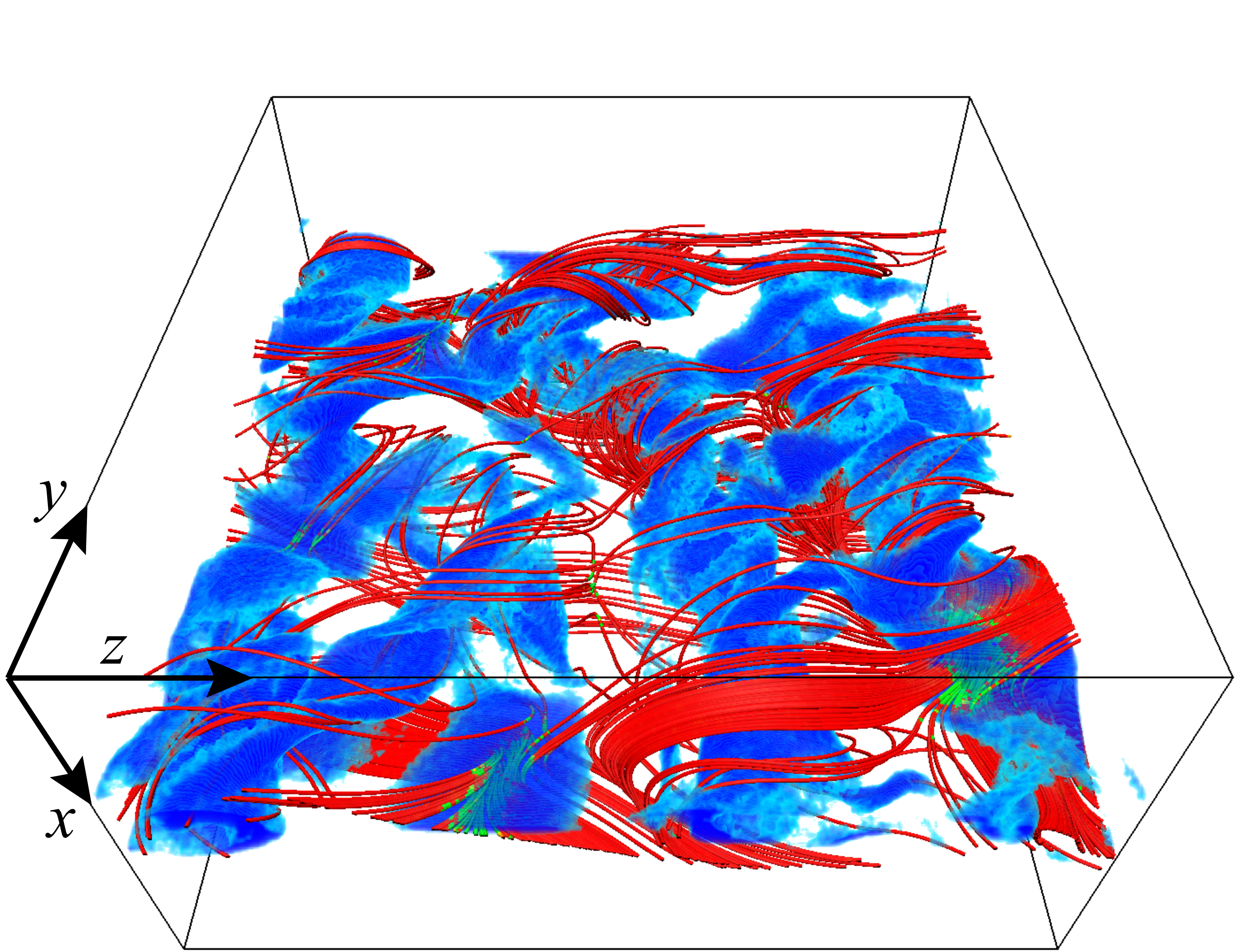}
\end{center}
\caption{A view of the current sheet at $x=L_x/4$ at times
  $t=56\,\tau_{c0}$ (upper panel) and $t=84.5\,\tau_{c0}$ (lower
  panel).  The volume rendering (blue color) shows the region with
  current density $J>0.2\,cB_0/(4\pi\lambda_{\rm p})$ (upper panel)
  and $J>0.12\,cB_0/(4\pi\lambda_{\rm p})$ (lower panel). The curves are the magnetic field lines that pass
  through regions of high parallel electric field $E_\parallel$ and thus are actively undergoing
  reconnection.  The fraction of the field line with an enhanced
  value of $E_\parallel$ is rendered in green. \label{fig:vapor}}
\end{figure}

The two rightmost columns of Figure \ref{fig:projections} show that the 
magnitude of the parallel electric field $E_\parallel \equiv |{\mathbf E}_\parallel|$ strongly
peaks in the thin, long current sheets containing primary X-lines and
connecting magnetic flux ropes. However prior to flux rope disruption and the
transition to a disordered state, the parallel field is normally
very small inside magnetic flux ropes even when they contain 
embedded, substructure-level current sheets.  An exception are flux
rope segments directly undergoing
dynamical transformation, such as merging or stretching, where very
isolated, spatially and temporally intermittent sites of significant
$E_\parallel$ 
occasionally appear. The flux rope interiors are
not undergoing pervasive, steady reconnection, in spite of their
complex magnetic substructure.  The intermittent $E_\parallel$ is also common
after the flux rope network has become disordered, indicating that the
disordered network remains active by developing
transient, localized dissipation events in the braided network of 
disrupted flux ropes that effect further evolution of magnetic
connectivity.  

Reconnection intermittency is also evident in Figure \ref{fig:vapor}, showing
  views of the current sheets $t=56\,\tau_{c0}$,
during the initial flux rope merging phase, and at
$t=84.5\,\tau_{c0}$, after the flux rope has become disordered.  The
field lines shown are those passing through the sites where
$E_\parallel$ is significant, where the field line color is shown in
green and field line geometry is that of an X-line.  The reconnected
magnetic field snaps back to become incorporated into that of the flux
ropes.  It is also noticeable that the flux ropes are not
cylindrically symmetric but highly flattened (ribbon-like) and exhibit a clear longitudinal twist.

A common measure of magnetic connectivity is the concept of magnetic
helicity, but its definition on the entire periodic domain presents unique challenges
  \citep[e.g.,][and references therein]{berger_magnetic_1997}.  In spatially and
  temporally localized, simply-connected
  reconnection regions, a generalized helicity can be meaningfully
defined using the Finn-Antonsen approach \citep[e.g.,][]{berger_introduction_1999}. In guide-field reconnection, a
change of the generalized helicity implies a change of global field
line connectivity \citep{schindler_general_1988}.
Since the generalized helicity is locally generated by
the source term $-2\, {\mathbf E}\cdot {\mathbf B}=-2\,{\mathbf
  R}\cdot{\mathbf B}$ that is closely related to
the parallel electric field, the twisting and braiding of the field
lines interior to a flux rope is a consequence of (possibly 
$y$-coordinate dependent) field connectivity
change in the primary X-line region.\footnote{The sign
of ${\mathbf E}\cdot{\mathbf B}$ should be uniform in each individual
X-line region.}

\subsection{Instability Mode Analysis}
\label{sec:mode_analysis}

\begin{figure}
\begin{center}
\includegraphics[width = 0.45\textwidth]{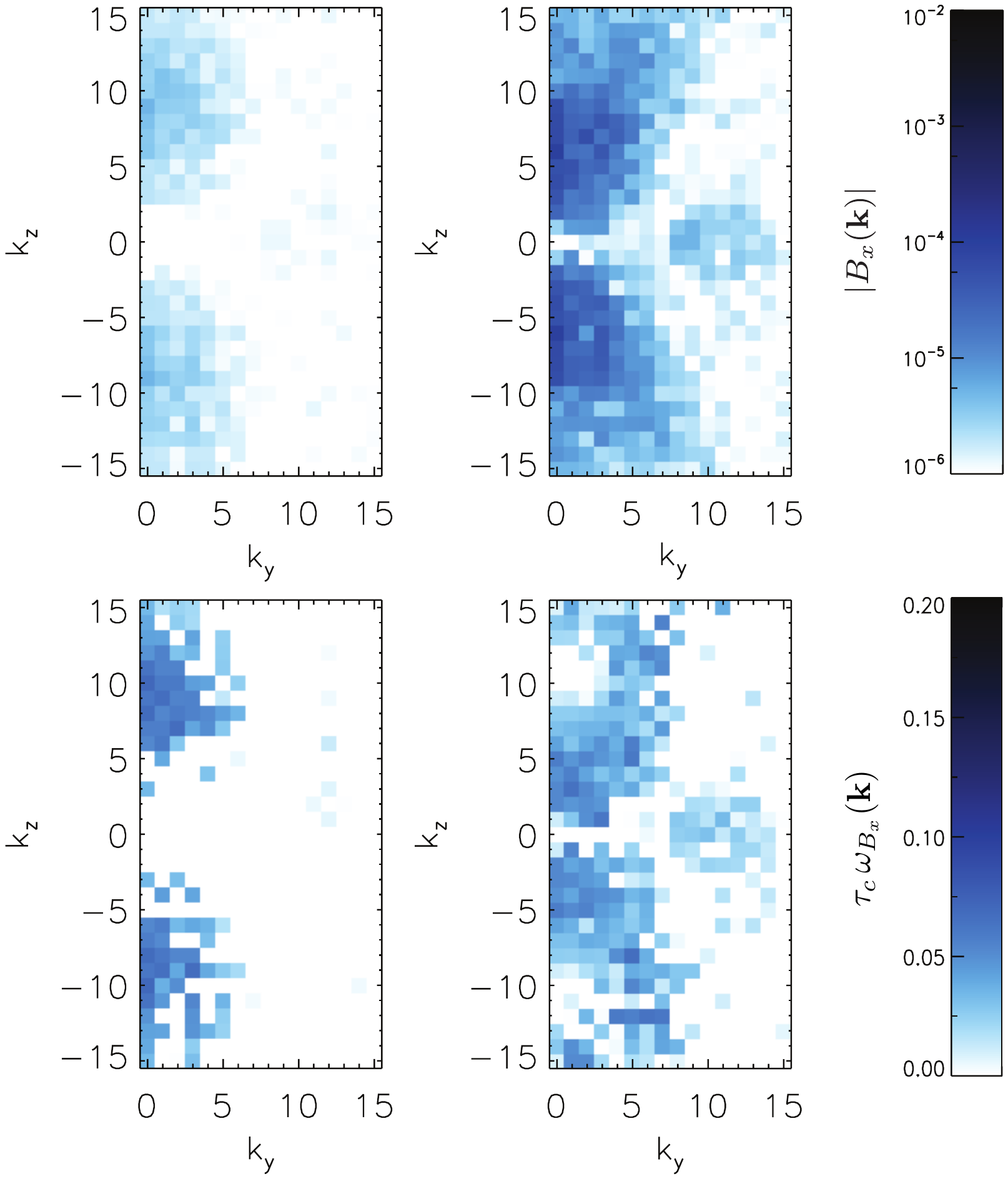}
\caption{Fourier mode amplitudes (top panels) and growth rates (bottom
  panels) of $B_x$ in one of
  the current sheets at times $t=19\,\tau_{c0}$ (left panels) and
  $t=37.5\,\tau_{c0}$ (right panels) in simulation {\tt S2K025L}.  The
  amplitude was computed with the
  Fast Fourier Transform (FFT) via ${\mathbf B}({\mathbf k})=(N_xN_yN_z)^{-1}
  \sum_{i_x=1}^{N_x} \sum_{i_y=1}^{N_y} \sum_{i_z=1}^{N_z} \exp[-2\pi
  i (i_y k_y /N_y+ i_z k_z/N_z) ] \,{\mathbf B} (i_x, i_y, i_z)$, where $N_x$,
  $N_y$, and $N_z$ are dimensions of the computational grid, and the
  initial reversing field $B_0=0.1$ in these units.
The components of the
  wave vector are expressed in units of $2\pi$ divided by
  the computational box size along the relevant direction.  We only
  show growth rates for modes with amplitudes $|B_x({\mathbf k})| > 10^{-6}$.
\label{fig:fft}}
\end{center}
\end{figure}

To identify modes present in the simulation, we carried out Fourier
decomposition of the electric and magnetic field for wave vectors
${\mathbf k}$
parallel to the plane of the current sheet, $k_x=0$, in simulation {\tt S2K025L}.  We search for signatures of
the tearing and kink instability, as well as for oblique modes, see, e.g.,
\citet{daughton_role_2011} and \citet{baalrud_reduced_2012}.  The
tearing instability
has ${\mathbf k}\parallel \hat{\mathbf z}$ and creates a perturbation in $B_x$ and $E_y$.  The kink instability has ${\mathbf k}\parallel
\hat{\mathbf y}$ and in the linear order creates a perturbation in
$E_z$ if a guide field is present.
The oblique mode is similar to the tearing mode, except that the wave vector is
tilted in the $yz$ plane, that of the initial current sheet.  In
Figure \ref{fig:fft}, we show Fourier power spectra of the magnetic
field component $B_x$ in
the linear phase at $t=19\,\tau_{c0}$ and at the end of the fully
nonlinear flux rope merging
phase at  $t=37.5\,\tau_{c0}$.   We also show the corresponding growth
rates evaluated over an interval of $\Delta t = 9\,\tau_{c0}$ preceding
each of the two times,
\begin{equation}
\omega_{B_x} ({\bf k},t) \equiv \frac{\ln|B_x({\mathbf k},t)/B_x({\mathbf
    k},t-\Delta t)|}{\Delta t} .
\end{equation}
In the figure, the growth rates are multiplied by the light crossing
time of the compressed current sheet $\tau_c\approx
0.43\,\tau_{c0}$.  The modes with fastest growth rates are clustered
around the $k_z=\pm2\pi (L_z/9)^{-1}$ tearing mode at $t=19\,\tau_{c0}$,
but the fastest growth then shifts to longer wavelengths and oblique
directions by $t=37.5\,\tau_{c0}$.  

At the end of the linear phase, 
the power spectrum for $B_x$ and $E_y$ peaks at
$(2\pi)^{-1}(k_y,k_z)=(0,\pm(L_z/9)^{-1})$, which is the tearing mode, consistent with
the presence of $\sim 9$ flux ropes in each current sheet.  The peak,
however, is broadened, likely by an initial lack of
phase coherence on scales given by Equation (\ref{eq:coherence_length}).  For the tearing
mode, $\omega_{\rm KTI}\sim 0.083\,\tau_c^{-1}$.
From this we expect lack of phase
coherence on scales $\Delta y,\, \Delta z\gg \lambda_{\rm coh,KTI}\sim
5.2\,\lambda_0\sim 16\lambda_{\rm p}$, which
can explain the observed broadening.  However the broadening also
allows the possibility that
an authentic oblique component is present, too. At the end of the flux
rope merging phase, which we define as the time $t\sim (50-60)\,\tau_{c0}$
when the orderly flux rope merging gives way to a more disordered
evolution,
 the peak has shifted to longer wavelengths $(L_z/2)^{-1} \leq (2\pi)^{-1} |k_z|
\leq (L_z/6)^{-1}$ with significant additional power occurring in the oblique
direction, for $(2\pi)^{-1} k_y\leq (L_y/3)^{-1}$.  We emphasize, again,
that the oblique power may be an outcome of phase decoherence, as
well as of secondary instabilities developing in the nonlinear regime, 
rather than of a genuine primary oblique instability mode.

Turning to our search for the kink mode, we find that $E_z$ develops
nonvanishing albeit weak ${\mathbf k} \parallel \hat{\mathbf y}$ power at
$t\gtrsim 40\,\tau_{c0}$ with the peak wavelength corresponding to
$(2\pi)^{-1} k_y = (L_y/3)^{-1}$.  However, most of the power in $E_z$
is still along ${\mathbf k} \parallel \hat{\mathbf z}$. Other
components of ${\mathbf E}$ and ${\mathbf B}$ are devoid of power
along  ${\mathbf k} \parallel \hat{\mathbf y}$.  This suggests that
the simulation exhibits no evidence for a linear kink mode and that
any variation with $y$ emerges in the nonlinear development of the
tearing mode.  We compare these results to previous analyses of
tearing, kink, and oblique modes in Section \ref{sec:mode_discussion} below.

\subsection{Flux Rope Merger Timescales}
\label{sec:timescales}

Once the tearing instability has produced a collection of
nonlinear flux ropes at a time which we will denote with $\tau_{\rm NL}$, the ropes begin to merge.  The merging time
should scale with the island separation $\Delta z$ as 
\begin{equation}
\label{eq:tau_merge}
\tau_{\rm merge} \sim \frac{\Delta z}{\chi\,v_{\rm A}} ,
\end{equation}
where $\chi$ is a dimensionless coefficient encapsulating the dynamics
of interaction between flux ropes.   
Then, the number of flux ropes per unit length along
the current sheet, which we denote with $\mu$, can be estimated as
\begin{equation}
\label{eq:island_density_model}
\mu \sim \frac{1}{\chi\,v_{\rm A}\,(t - \tau_{\rm NL})} .
\end{equation}
 By visually counting the number of flux ropes in each
current sheet in simulation {\tt S2K025L} in Figure
\ref{fig:projections}, we found that
\begin{equation}
\label{eq:island_density_fit}
\mu \approx \frac{1.2}{\lambda_{\rm p}} \left(\frac{t}{\tau_{c0}} -
  16\right)^{-1} ,
\end{equation}
suggesting that $\tau_{\rm NL}\approx 16\,\tau_{c0}$ in that simulation.
Comparing equations (\ref{eq:island_density_model}) and
(\ref{eq:island_density_fit}) we estimate that $\chi\approx 0.4$.
Given that the reconnection outflow velocities reach only about a half
of the Alfv\'en velocity in the simulation, 
it is not surprising that the characteristic 
velocity associated with flux rope merging $\chi\,v_{\rm A}$  is only a
fraction of the Alfv\'en velocity. 

In reconnection configurations starting from a one-dimensional
current sheet lacking any $y$ and $z$ dependence, the
three-dimensionality manifested in an emerging $y$ dependence can
arise from several effects which include an intrinsic obliqueness of the
tearing modes, the development of kink modes, and a lack of coherence on scales
on which the phase of the tearing mode is uncorrelated  (see Section
\ref{sec:unstable_modes}).  In reconnection regimes in which oblique and kink modes are
suppressed, flux rope dynamics acquires three dimensional
character when the flux rope separation $\sim \mu^{-1}$ starts
substantially exceeding the
coherence length $\lambda_{\rm coh,KTI}$.  From equations
(\ref{eq:KTI_max}), (\ref{eq:coherence_length}), and
(\ref{eq:island_density_model}), this happens at times
\begin{equation}
t\gg \tau_{\rm 3D,coh} \sim \tau_{\rm NL} + \frac{\lambda}{0.35\,\chi\,v_{\rm
    A}\,\beta^{3/2}} ,
\end{equation}
yielding an estimate $\tau_{\rm 3D,coh} \sim36\,\tau_{c0}$ for
simulations {\tt S2K025} and {\tt S2K025L}.  Indeed, the flux rope geometry in Figure
\ref{fig:projections} seems to acquire three-dimensional
character at $t\gtrsim(40-50)\,\tau_{c0}$.  This is consistent with
the rise of 
power at oblique wave vectors $k_y
>0$ in Figure \ref{fig:fft} at similar times.  This analysis suggests that the reconnection layers
acquire three dimensional structure on relatively short time scales
regardless of the growth rates of the oblique and kink modes.

\subsection{The Reconnection Rate}
\label{sec:reconnection_rate}

We now discuss the reconnection efficiency at primary X-lines in the
early stages of the evolution of the current sheet.  In this regime,
the reconnection site can be described using a typical two-dimensional
model of reconnection in which two oppositely
oriented fields are separated by a thin current layer in which the
field reverses. In this layer, magnetic field lines diffuse across the
plasma to reconnect at one or more X-lines. Magnetized plasma
approaches the central plane of the layer, toward an X-line, with an
asymptotic inflow velocity $v_{\rm in}$, which is also known as the
reconnection velocity. After passing the X-line, plasma is expelled
from the vicinity of the X-line to either side at the outflow velocity
$v_{\rm out}$, which is typically assumed to equal the Alfv\' en velocity
$v_{\rm A}$.  The orientation of the reconnected magnetic field is
approximately perpendicular to that of the original field.
 
Defining $\delta$ and $L$ to be the thickness and length of the current sheet, the conservation of mass from the inflow to the outflow requires
\begin{equation}
\frac{\delta}{L}\approx \frac{v_{\rm in}}{v_{\rm out}} \approx \frac{v_{\rm in}}{v_{\rm A}} .
\label{sprec}
\end{equation}
The inflow velocity can be related to the electric field in the
reconnection region $\mathbf{E}_{\rm RR}$ which is approximately
uniform and equal to the dynamical electric field the plasma inflow
just outside the current sheet
\begin{equation}
{\mathbf E}_{\rm RR}\approx \frac{1}{c} ({\mathbf v}_{\rm in}\times {\mathbf B_0}) .
\label{eq:estructure}
\end{equation}
Because the inflow velocity and the reversing magnetic field are
perpendicular, the dimensionless reconnection rate $r_{\rm rec}$,
defined as the ratio of the inflow and the outflow velocity, and the electric field in the reconnection region, can be related via
\begin{equation}
\label{recE}
r_{\rm rec}\equiv \frac{v_{\rm in}}{v_{\rm out}}=\frac{E_{\rm RR}}{(v_{\rm A}/c)B_0} .
\end{equation}
This calculation of the reconnection rate makes the common assumption
that the outflow velocity is equal to the Alfv\'en velocity in the inflow
region.  In our simulations, this assumption is not valid as the
outflow velocity is smaller than the Alfv\' en velocity.

In simulation {\tt S2K025}, we find
that indeed, the total electric field  in the reconnection region is
approximately uniform across the current sheet. Given that the
Alfv\'en velocity is $v_{\rm
  A}\approx 0.7\,c$ in this simulation, the dimensionless
reconnection rate calculated from the electric field following
Equation (\ref{recE}) is $r_{\rm
  rec}\approx0.05$ at $t=37.5\,\tau_{c0}$, in the middle of the initial flux rope
merging phase. This is similar to the peak reconnection rates of
about $0.07-0.1$ found in other three dimensional PIC simulations 
with higher magnetizations
\citep[e.g.,][]{zenitani_role_2008,liu_particle_2011}. 
This comparison 
 assumes that reconnection at the largest X-line can be
described accurately by a two dimensional reconnection model. The inflow and outflow
velocities at $t=37.5\,\tau_{c0}$ are $v_{\rm in}\approx0.02\,c$ and
$v_{\rm out}\approx0.27\,c$, respectively. The peak plasma outflow
velocity slightly later in the simulation reaches $v_{\rm max}\approx
0.39\,c$, just above half of the Alfv\'en velocity. Since the
Alfv\'en velocity is not ultrarelativistic, it is not surprising that
we are not finding ultrarelativistic flows in any of our simulations.
 \footnote{
To obtain an ultrarelativistic flow, one must 
have $\gamma_{\rm A}\equiv[1-(v_{\rm A}/c)^2]^{-1/2}\gg 1$, but this
may not be
sufficient, e.g., \citet{zenitani_role_2008} found maximum flow speeds of $v_{\rm out}\approx (0.7-0.8)\,c$ even in their two-dimensional simulations with high magnetization and no guide field.}  If the
dimensionless reconnection rate is calculated directly from the inflow
and outflow velocities, it is $r_{\rm rec}\approx 0.08$.  The difference between the
outflow velocity and the Alfv\'en velocity, which are equal in
idealized two dimensional reconnection models, is likely due to the dynamical,
non-steady state nature of the plasma flow in our simulations; similar discrepancies were found by  \citet{bessho_fast_2010} in their nonrelativistic PIC simulations, and \citet{takahashi_scaling_2011} in their relativistic MHD simulations.

\subsection{Generalized Ohm's Law Analysis}
\label{sec:ohms_law}

\begin{figure}
	\centering
\includegraphics[width=0.475\textwidth]{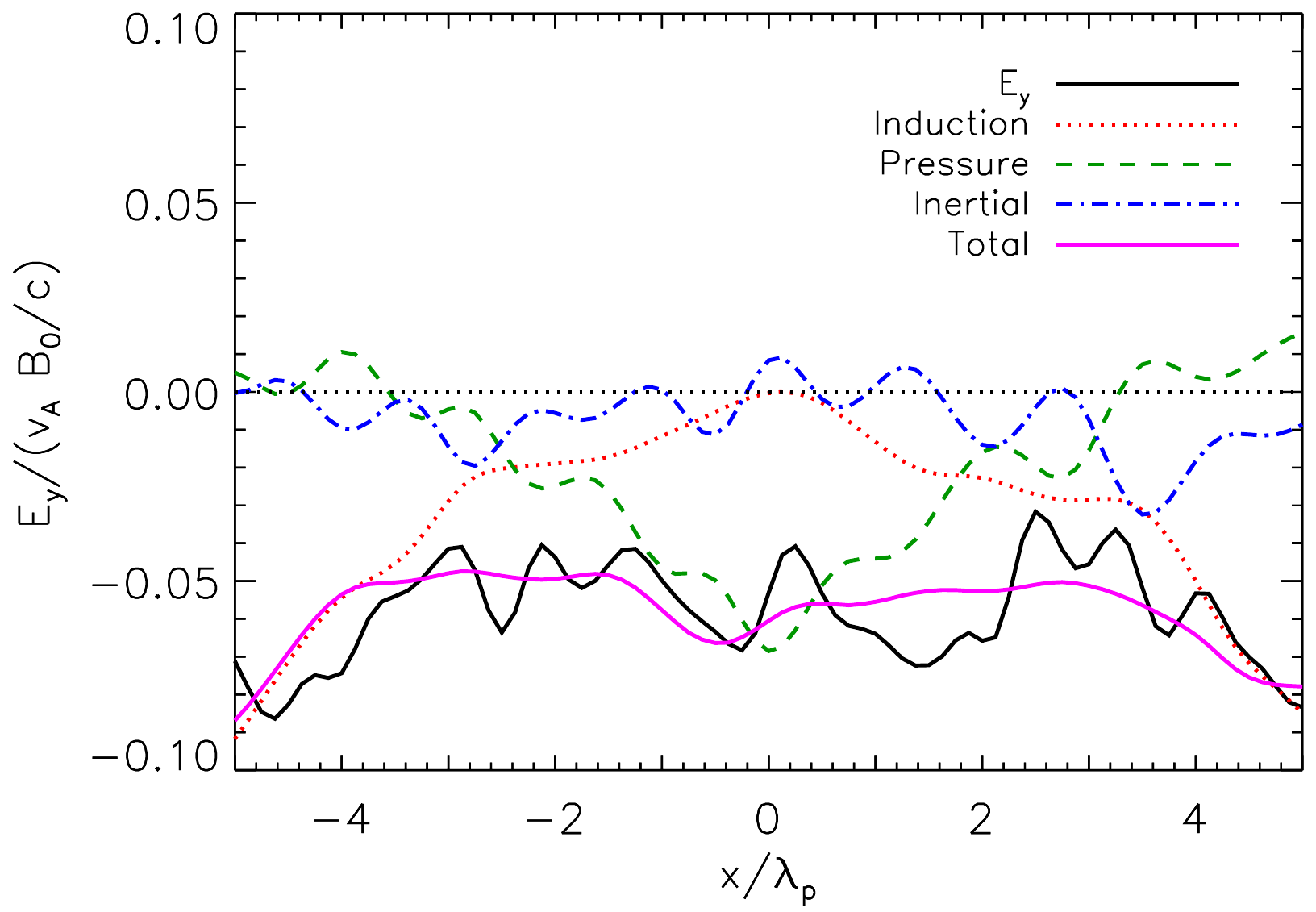}
	\caption{The variation of the terms in the generalized Ohm's law 
          of the electric field across the current sheet near the
          largest X-line at $t=47\,\tau_{c0}$. The center of the current
          sheet is at $x=0$. The lines represent the measured electric
          field $E_y$ (solid, black), the induction
          component (dotted, red), the pressure component (dashed,
          green), the inertial component (dot-dashed, blue), and the
          total (pink, solid) of the three calculated electric field
          terms. The generalized Ohm's law mandates that the latter
          should equal the actual electric field, which is the case
          apart from fluctuations arising from numerical
          discreteness. The Ohm's law components are smoothed with a
          Gaussian kernel of standard deviation $0.33 \lambda_{\rm
            p}$. It is clear that at the X-line, the pressure term
          dominates over the inertial term.\label{fig:ohmslawy}}
\end{figure}

A central question in magnetic reconnection is the nature of the process
that facilitates the violation of flux freezing in the diffusion region. 
The total electric field ${\mathbf E}$ in a collisionless plasma is
related to the magnetic field ${\mathbf B}$ and to moments of the
particle distribution function of particle species $s$ by the generalized Ohm's law \citep[e.g.,][]{krall_principles_1973}
\begin{eqnarray}
 {\mathbf E} + \frac{1}{c}\langle {\mathbf v}_s\rangle \times {\mathbf B}&=&
        \frac{1}{q_s n_s} \nabla \cdot {\mathbf P_s}\nonumber\\& &+\frac{m_s}{q_s} \left(\frac{\partial \langle {\mathbf p_s}\rangle}{\partial t} + \langle {\mathbf v_s} \rangle \cdot \nabla \langle {\mathbf p_s}\rangle\right) ,
\label{eq:g_ohms}
\end{eqnarray} 
where $\langle {\mathbf v}_s \rangle$ and $\langle {\mathbf p}_s \rangle$
are the average velocities and momenta of particles, ${{\mathbf P}_s}$
is the pressure tensor, $n_s$ is the particle number density, and $q_s$
and $m_s$ are the particle charge and mass. Averaging over the two
species $s=\{i,e\}$ we recover the non-ideal, reconnection electric
field ${\mathbf R}$ on the left hand
side of Equation (\ref{eq:g_ohms}).  In collisionless plasmas with long-lived magnetization
(counterexamples being the plasmas excited by kinetic instabilities,
such as the filamentation instability), the pressure and the inertial
terms on the right hand side of Equation (\ref{eq:g_ohms}) are small
almost everywhere. Both of these terms, however, can become important
in the diffusion region of a magnetic reconnection layer.

To determine which terms give rise to ${\mathbf R}$, we calculate the terms on both sides of
Equation (\ref{eq:g_ohms}) as they vary across the X-line in simulation
{\tt S2K025}. We find no significant
reconnection electric field in the $x$ or $z$ directions; there
is only a reconnection field in the $y$ direction, consistent 
with two dimensional models of reconnection.  Meanwhile, during the
linear growth phase of the tearing instability, spatial gradients are
present only in the $x$ and $z$ directions, implying that the
off-diagonal $xy$ and $zy$ components of the pressure tensor
contribute to the first term on the right hand side of Equation
(\ref{eq:g_ohms}).  In particular, we find that most of the pressure
term is provided by $\partial P_{xy}/\partial x$. Figure
\ref{fig:ohmslawy} shows the $y$ component of various terms
 contributing to the electric field. The pressure term is
dominant, and the total of the inductive, pressure, and inertial terms approximates the 
electric field, with departures arising from shot noise.

These results are consistent with previous
investigations
\citep[e.g.,][]{bessho_collisionless_2005,bessho_fast_2007,schmitz_kinetic_2006} that
found that the spatial variation of off-diagonal terms of the pressure tensor was 
the main contributor to the reconnection
electric field. 
The results of our Ohm's law calculation with a moderate guide field,
combined with the similar results of other simulations with strong
guide field \citep[][nonrelativistic simulations]{che_current_2011} and no guide field
\citep[][relativistic simulations]{liu_particle_2011}, constitute evidence that for moderate
magnetizations and all guide field magnitudes, the reconnection
electric field is produced by the pressure term in three as well as
in two dimensions.  
In an exception to this agreement, \citet{hesse_dissipation_2007} found that the inertial term
became important in highly magnetized relativistic current sheets
with a guide field, which is a regime we do not probe.

\subsection{Particle Energization}
\label{sec:energization}

\begin{figure}
\includegraphics[width = 0.45\textwidth]{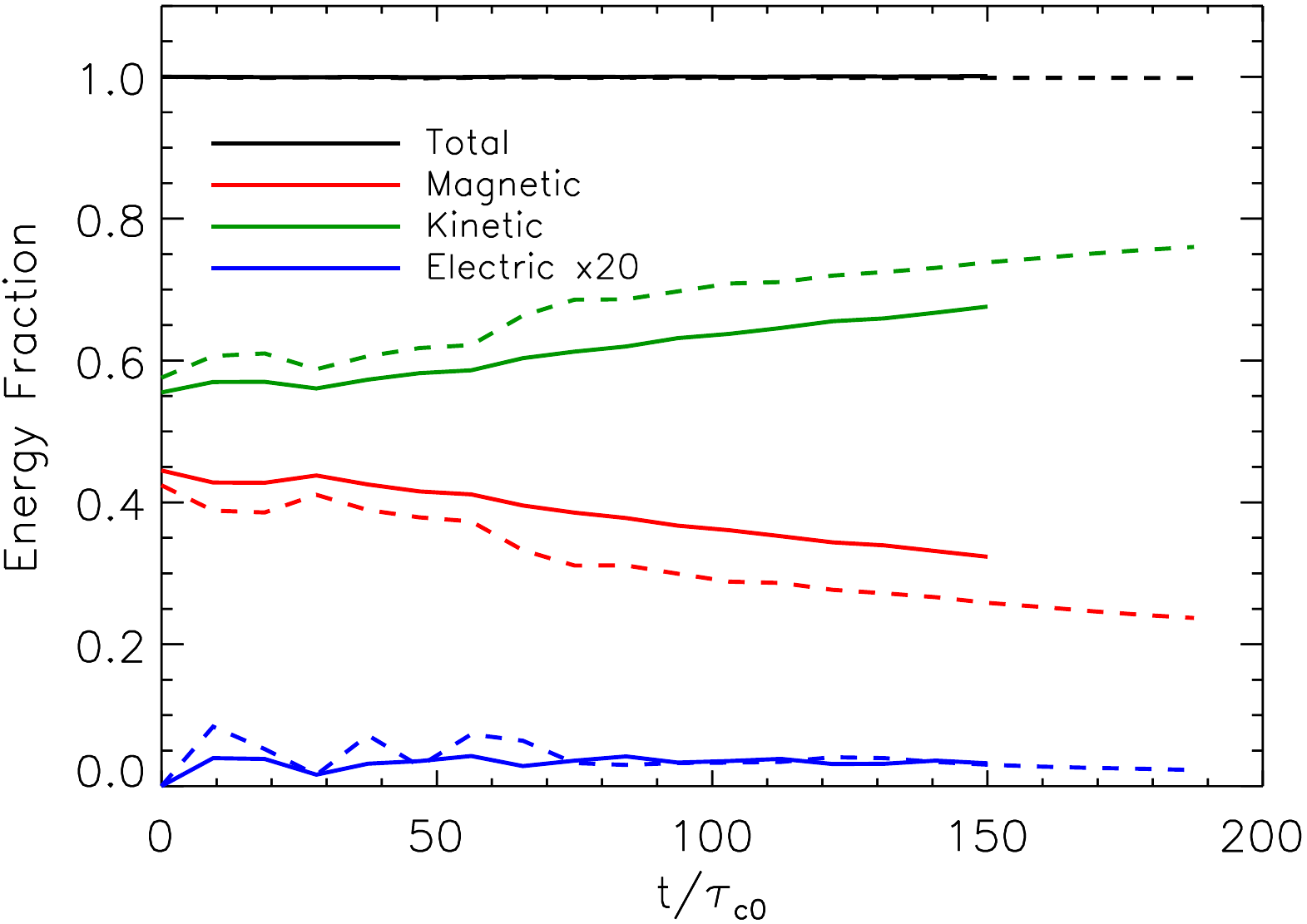}
\caption{Evolution of the total magnetic energy (red lines),
  particle kinetic energy (green lines), and the energy in the
  electric field multiplied by $20$ for clarity 
(blue lines), all normalized to the total initial
  energy, in runs {\tt S2K025L} (solid lines) and {\tt S2K025} (dashed lines).
  The time is expressed in units of the light crossing time of the
  current sheet $\tau_{c0}$.\label{fig:energy_evolution}}
\end{figure}

We turn to the topic of magnetic to kinetic energy conversion and
particle energization.
Figure \ref{fig:energy_evolution} shows the evolution of the total
magnetic field, electric field, and particle kinetic energies in
simulations  {\tt S2K025}  and {\tt S2K025L}.  
Energy was conserved to within $0.3\%$ in all the
simulations.  The initial kinetic and electromagnetic energy fraction
differs in the two simulations because the initial current sheet
occupies a fraction of the volume of the simulation box that is twice
as large in
simulation {\tt S2K025}  as in {\tt S2K025L}.  The trends, however,
are consistent taking the volume difference into account. 
An initial reversible fluctuation of the magnetic-to-kinetic ratio during the
first $\sim 20\,\tau_{c0}$ is associated with the readjustment to pressure equilibrium.
Dissipative conversion of magnetic to kinetic energy 
seems to first occur in a couple of bursts at $t\sim(30-40)\,\tau_{c0}$ and
$t\sim(50-60)\,\tau_{c0}$.  
Then, the system enters a phase, lasting until the end of
the simulation, in which dissipation is relatively steady.  The two bursts seem to
coincide with initial tearing instability and the formation of $\sim9$ flux ropes (magnetic
islands) across the simulation box per current sheet, and the subsequent flux rope
merging, yielding $\sim 3$ flux ropes per current sheet by $t\sim60\,\tau_{c0}$, after which
the overall geometry of the reconnection layer becomes fully three
dimensional.  By
$t=150\,\tau_{c0}$, about $40\%$ and $20\%$ of the initial magnetic energy is
converted into particle kinetic energy in the smaller and larger
simulation, respectively.  

In Table \ref{tab:simulations}, we show $|\Delta
{\mathcal E}_B|/{\mathcal E}_B$, the fraction of magnetic field
energy that is converted to particle kinetic energy in each
simulation. Here, we define ${\mathcal E}_B$ to be the initial energy in the total magnetic field, which
includes both the reversing field and the guide field, and
$\Delta {\mathcal E}_B$ is the change of the total magnetic energy from the beginning to the end of
the simulation.  The
energy in the reversing field only is a factor of $\approx (1+\kappa^2)^{-1}$
smaller than ${\mathcal E}_B$ if we ignore the reduction of the reversing field in the
current sheets.  Independent of this correction, $|\Delta
{\mathcal E}_B|/{\mathcal E}_B$ exhibits a strong decreasing trend with
increasing guide field strength $\kappa$.  It also exhibits a weaker increasing trend
with $\sigma$; note that the very high $|\Delta {\mathcal E}_B|/{\mathcal E}_B$ in simulation
{\tt S1K0} is the result of destructive current sheet interaction.   
In the simulations with $\kappa\geq 0.5$, and also in simulation {\tt S1K025}
with $\kappa=0.25$  and $\sigma=1$, the conversion of magnetic to
kinetic energy is substantially weaker than in the other
simulations. We  also note that the rate of development and evolution
of the flux rope network
exhibits similar behavior in the sense that flux ropes of a given size
assemble later ($\sim 80\,\tau_{c0}$ vs.\  $\sim40\,\tau_{c0}$) in the
$\kappa\geq 0.5$ and {\tt S1K025} simulations.

The variation of $|\Delta
{\mathcal E}_B|/{\mathcal E}_B$ and of the overall evolution rate with $\kappa$
is likely a
consequence of the decrease of plasma compressibility with increasing
guide field strength \citep[e.g.,][and references
therein]{zenitani_self-regulation_2008}.  Since most of the magnetic
to particle kinetic energy conversion takes place during the
relaxation of reconnection field lines to their final equilibrium in
flux ropes, the less compressible case of a strong guide field leads
to larger flux ropes and smaller energy conversion.

\subsubsection{Energization Efficiencies}
\label{sec:energization_efficiencies}

\begin{figure}
\includegraphics[width = 0.475\textwidth]{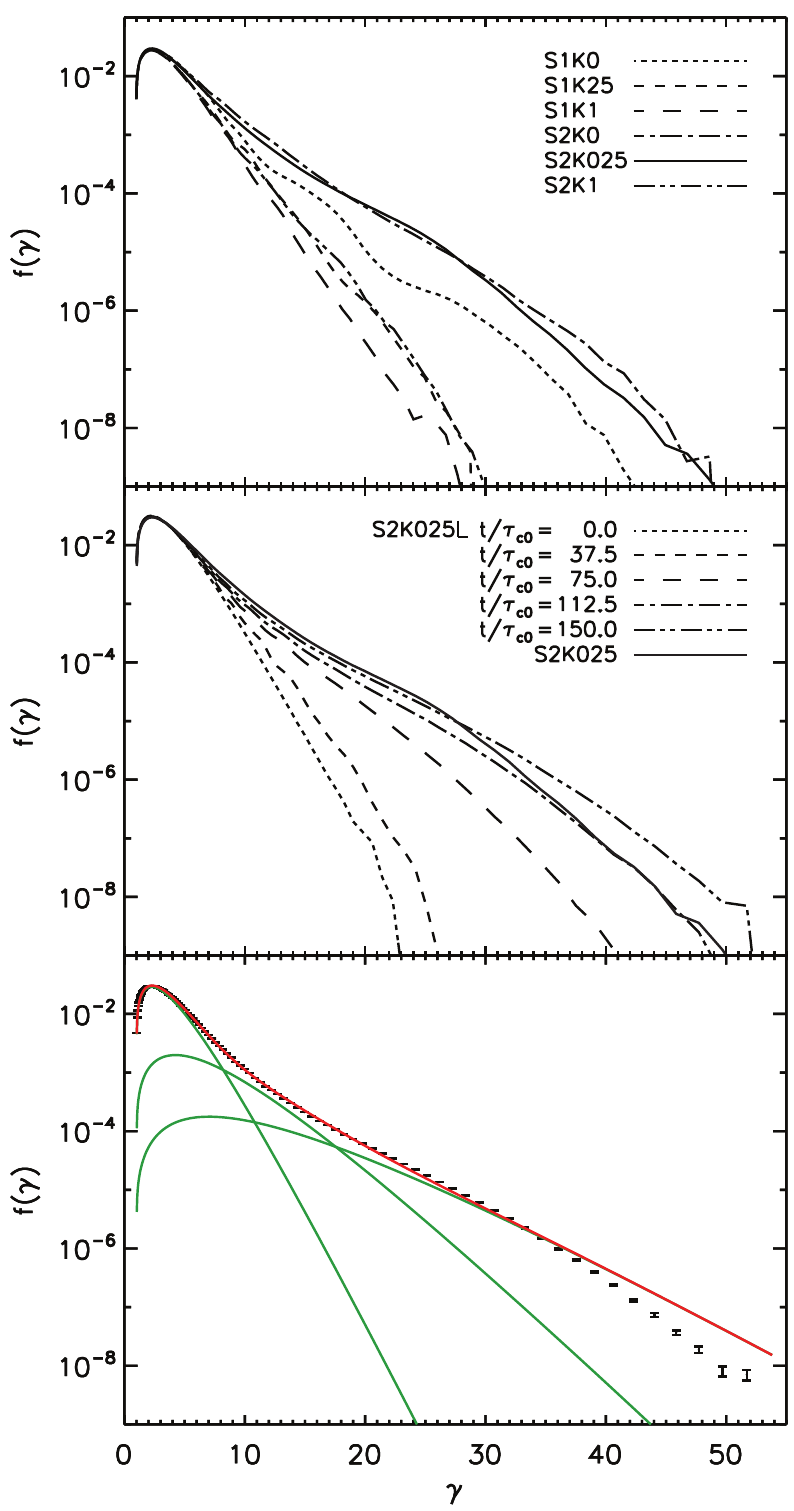}
\caption{Particle energy spectra in the six smaller size simulations
  at time $t=150\,\tau_{c0}$ (top panel; see legend) and in the large size 
simulation {\tt S2K025L} at five different times (middle plan; see legend) where for reference
we include the $t=150\,\tau_{c0}$
  spectrum from the corresponding smaller size simulation {\tt S2K025}
  (solid line). The bottom panel compares the final spectrum to a
  a model containing three thermal populations at different
  temperatures (red line; see text), including the spectra of the
  three individual thermal components (green lines).  The spectra and
  uncertainties are calculated from a random sample containing $5\%$
  of the particles in the simulation.\label{fig:energy_spectrum}}
\end{figure}

Figure \ref{fig:energy_spectrum}, top panel, shows the energy spectrum
at a reference time $t=150\,\tau_{c0}$ in six of the smaller size simulations.  Each of
the spectra exhibits a peak in agreement with the initial thermal
distribution as well as a new tail resulting from particle energization 
extending to maximum Lorentz factors in
the range $\gamma_{\rm max}\sim 30-50$ (for the precise values at the
end of the simulations, see
Table \ref{tab:simulations}).  The simulations with the strongest particle
energization are those with higher magnetization $\sigma=2$ and zero
or moderate guide field $\kappa\leq 0.25$.  Rather weak energization
is seen in all simulations with strong guide field $\kappa=1$  and in
the simulation {\tt S1K025} with weak magnetization and moderate guide
field. We observe an intermediate level of energization in simulation
{\tt S1K0} with weak magnetization and no guide field; this simulation
is unique in that at the reference time, the two current sheets have
already interacted with each other.  It is important to note that all
of these spectra are calculated at the same reference time; therefore,
these differences in energization efficiency might result from the
differing energy conversion rates discussed in the previous section.
However, we will find below that the spectrum exhibits the strongest
evolution during the flux rope merging phase; because the merging
phase is complete by the reference time in all the simulations, it
seems that the differences in the level of energization between the
runs cannot be attributed solely to the differences in the rate of reconnection.

In order to quantify the degree of particle energization in the
simulations, 
following \citet{zenitani_particle_2007}, we compute the particle
kinetic energy $K_{\rm ener}$
contained in the difference between the measured spectrum and a thermal spectrum
containing the same number of particles and the same energy at
particle energies at which the measured spectrum is in excess of the
thermal spectrum.  We then calculate the ratio of $K_{\rm ener}$ to the
total particle kinetic energy $K$, both computed at the end of the
simulation.  It is worth remarking that we do \emph{not} 
go as far as \citet{zenitani_particle_2007,zenitani_role_2008} to identify the
particles contributing to $K_{\rm ener}$ as a nonthermal population;
the following section we show this tail may be better described with a
combination of several thermal populations.
Consistent with the trend observed in the shape of the spectrum, 
we find that $K_{\rm ener}/K$, which is shown in Table
\ref{tab:simulations}, 
is the largest in the simulations with the higher magnetization $\sigma=2$
and at most moderate guide field $\kappa\leq 0.25$ and is the
smallest in simulation {\tt S1K1} with lower magnetization $\sigma=1$ and
strong guide field $\kappa=1$.
Thus, in contrast with \citet{zenitani_role_2008}, we find that $K_{\rm ener}/K$ decreases with
increasing guide field.  The reason for this dissimilarity seems to be that the kink
instability does not compromise the development of reconnection in any
of our simulations other than {\tt S1K0}; even in that run, a minor
flux rope merger occurs before the current sheet is disrupted.  Without the
kink instability, more particle energization can take place.

Another possibly more interesting measure of particle energization is the ratio of the kinetic energy in
energized particles to the magnetic energy converted to kinetic
energy.  Combining equations (\ref{eq:harris_field}) and (\ref{eq:sigma_uniform}) with the definition of $\Gamma$, and taking account of the conservation of energy  $K(t)=K(0)+|\Delta {\mathcal E}_B|$, this ratio can be written as
\begin{equation}
\frac{K_{\rm ener}}{|\Delta {\mathcal E}_B|} =
\left(\xi\left|\frac{\Delta
      {\mathcal E}_B}{{\mathcal E}_B}\right|^{-1}+1\right)\frac{K_{\rm ener}}{K} .
\end{equation}
Here, the dimensionless coefficient $\xi$ is the ratio of the initial particle kinetic energy density to the magnetic energy density including the guide field, given by
\begin{equation}
\xi\approx \frac{1}{\sigma(1+\kappa^2)}
\left(\frac{1}{\Gamma-1}-\frac{1}{\theta_0}\right) ,
\end{equation}
where $\theta_0\equiv T_0/(m_e c^2)$ and all the quantities 
entering the definition of $\xi$ are evaluated at
the beginning of the simulation.  The ratio, which ranges between
$11\%$ and $38\%$, is shown in Table
\ref{tab:simulations}.  The trend seen at low magnetization $\sigma=1$ is an increase of $K_{\rm
  ener}/|\Delta {\mathcal E}_B|$ with guide field strength, approximately the opposite
of that seen in $K_{\rm ener}/K$. 
The trend can be understood by noting that, as we will find in Section \ref{sec:energization_mechanism},
most of the particle energization contributing to $K_{\rm ener}$
occurs in
X-line regions. Meanwhile, 
$\Delta {\mathcal E}_B$ measures the change in magnetic energy both
during reconnection in 
X-line regions and, more importantly, during the subsequent 
contraction of reconnected magnetic field lines
into flux ropes.  Therefore the ratio $K_{\rm ener}/|\Delta {\mathcal E}_B|$
should indeed increase with stronger guide field because the guide
field reduces plasma compressibility and moderates field line contraction.

\subsubsection{Energy Spectrum Evolution and Structure}
\label{sec:spectrum}

Figure \ref{fig:energy_spectrum}, middle panel, shows the evolution of the particle
energy spectrum in the large
simulation {\tt S2K025L}.  The spectrum evolves little during the
linear tearing phase and the first round of flux rope merging at
$t< 37.5\,\tau_{c0}$, but then it quickly hardens by $t=75\,\tau_{c0}$ as the flux rope
merging concludes and the reconnection layer transitions to
a disordered state.  Further hardening takes place until the end of the
simulation at $t=150\,\tau_{c0}$.  For comparison, we also show the
particle energy spectrum at this time from the smaller, equivalent
simulation {\tt S2K025}.  The two spectra are consistent at
Lorentz factors $\gamma \lesssim 28$ indicating convergence with
increasing box size at relatively low energies. However, the larger simulation has
progressively more particles at still higher energies, with Lorentz
factors reaching $\gamma\approx 50$. 

We proceed to model the entire
spectrum in simulation  {\tt S2K025L} as is, while keeping in mind
that in a still larger simulation, further evolution of the spectrum
at the highest energies is likely to be expected.  
We experimented with composite populations containing thermal as well
as power law components, and found that at Lorentz factors
$\gamma\lesssim 35$, a model containing three thermal populations,
each described by a relativistic Maxwellian, seems to work best.  The
model and the three thermal components are shown the bottom panel of
Figure \ref{fig:energy_spectrum}.
We fix the temperature of the first
population to be equal to the temperature of the initial plasma,
$T_1=m_e c^2$.
A least-squares fit yielded
temperatures $T_2=2.1\,m_ec^2$ and $T_3=3.5\,m_e c^2$ for the second
and third population, respectively.
The first component contains $85\%$ of the particles and
$70\%$ of particle kinetic energy, the second component
contains $13\%$ of particles and $24\%$ of particle kinetic energy, and
the third component contains only $2\%$ of the particles and $6\%$ of
particle kinetic energy.  We show in Section
\ref{sec:energization_sites} that the energized populations and
especially the particles with $\gamma > 30$ are
located close to the primary X-lines (within about one
Larmor radius) and in flux ropes.

\subsubsection{Energization Sites and Mechanism}
\label{sec:energization_mechanism}

 To gain insight in the nature of the particle energization process, 
in simulation {\tt S2K025}  we select a number,
 $N_{\rm trace} = 1841$, of particles reaching the highest energies,
 corresponding to Lorentz factors $\gamma\geq 32$, at the
 end of the flux rope merging phase at $t=75\,\tau_c$. We
 trace the orbits of these particles throughout the merging phase over the
 time interval $47\,\tau_c\leq t
 \leq 75\,\tau_c$. This is the period during which 
the particles experience coherent energization.  
The particles generally begin in the current sheet and have 
momenta that, on average, are aligned with the
direction of the electric force $q_j{\mathbf E}$ in the current sheet,
with a median inclination of $\approx 30^\circ$ from the direction of
the force. 

During flux rope merging, the traced particle energies increase approximately linearly.  
However at the instance of merging for the two largest flux ropes, which occurs at
$t\approx 61\,\tau_c$, most of the particles incur an energy increment
accounting for $\approx 15\%-20\%$ of the total energy gain.
Following the flux rope merging phase, some traced particles gain
additional energy, but others lose energy, and both the gain and the loss
could be considered manifestations of a thermalization process.  Indeed,
the energization $K_{\rm ener}/K$ does not further increase after the
flux rope merging is complete.

\label{sec:energization_sites}
\begin{figure}
\begin{center}
\includegraphics[width = 0.475\textwidth]{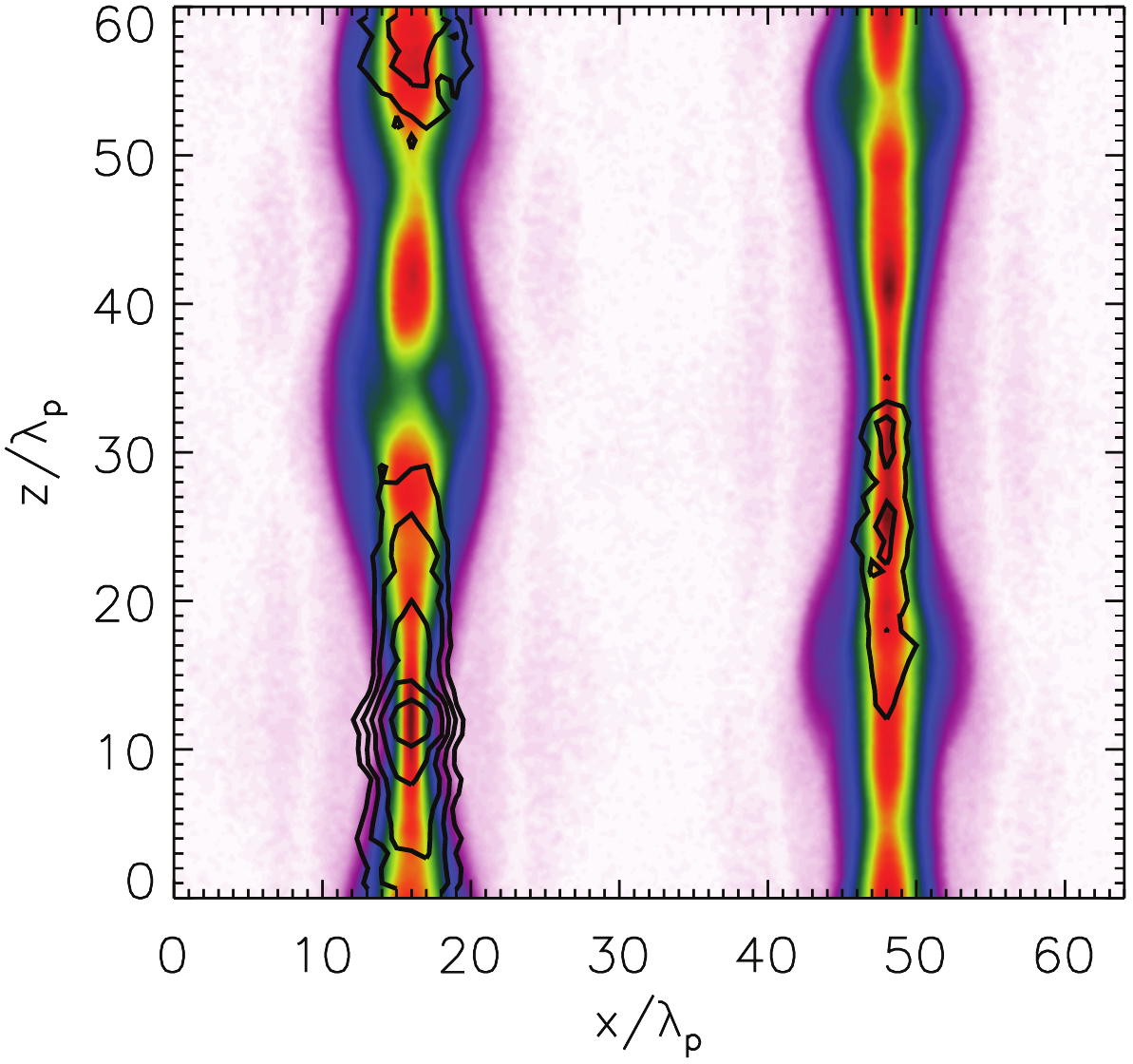}
\end{center}
\caption{The locus of particle acceleration in simulation {\tt S2K025}
  in the interval $47\,\tau_c\leq t\leq 70\,\tau_c$ corresponding to the primary
  particle energization phase.  The contours show the logarithm of the $y$-averaged
  energy gain per particle per unit volume $\Upsilon(x,z)$ for the
  $\approx 2000$ highest
  energy particles in the simulation in this interval (see text); the lowest
  contour corresponds to $0.04$ times the peak value of $\Upsilon$.  The underlying
  color image shows the projection in the $xz$ plane of the
  current density $J_y$ averaged over the interval; the current
  density is scaled linearly from light purple to dark red.
	\label{fig:acceleration}}
\end{figure}

To pin down the geometric location of particle energization, for the
traced particles, we
compute the total energy gain per particle per unit volume averaged
over $y$ via
\begin{eqnarray}
 \Upsilon (x,z) &=& \frac{1}{N_{\rm trace}}\sum_{j=1}^{N_{\rm trace}} \frac{q_j}{m_e c^2}
\int \mathbf{v}_j(t)\cdot \mathbf{E}[{\mathbf x}_j(t)]\nonumber\\
& &\times \delta[{\mathbf x}-{\mathbf x}_j(t)]\, dt \,\frac{dy}{L_y} ,
\end{eqnarray}
where ${\mathbf x}_i(t)$ and ${\mathbf v}_i(t)$ are the position and
velocity of traced particle $j$ at time $t$, 
$\delta$ is the Dirac delta, and the time integral
covers the flux rope merging phase.  In Figure \ref{fig:acceleration}, 
we compare the energy gain $\Upsilon$ to the corresponding absolute
value of the
$y$- and $t$-averaged current density $J_y$.  The highest energy gain
is clearly associated with the largest, best defined X-line on the
lower left.  Substantial energization is also detected in the outflow
regions flanking this X-line. The total energy gain in the
flanking regions, seen vertically above and below the X-line in the
figure, 
is larger than that in the narrow vicinity of the
X-line.  

It is puzzling that the
overall energization efficiency is weaker in the second current sheet shown on
the right in the figure. In both sheets, the energization seems to be
associated with some current density maxima but not with others.
Energization seems to prefer long, continuous X-line regions with
thin current sheets, perhaps because these are the sites of plasma
inflow.

\begin{figure}
\begin{center}
\includegraphics[width=0.43\textwidth]{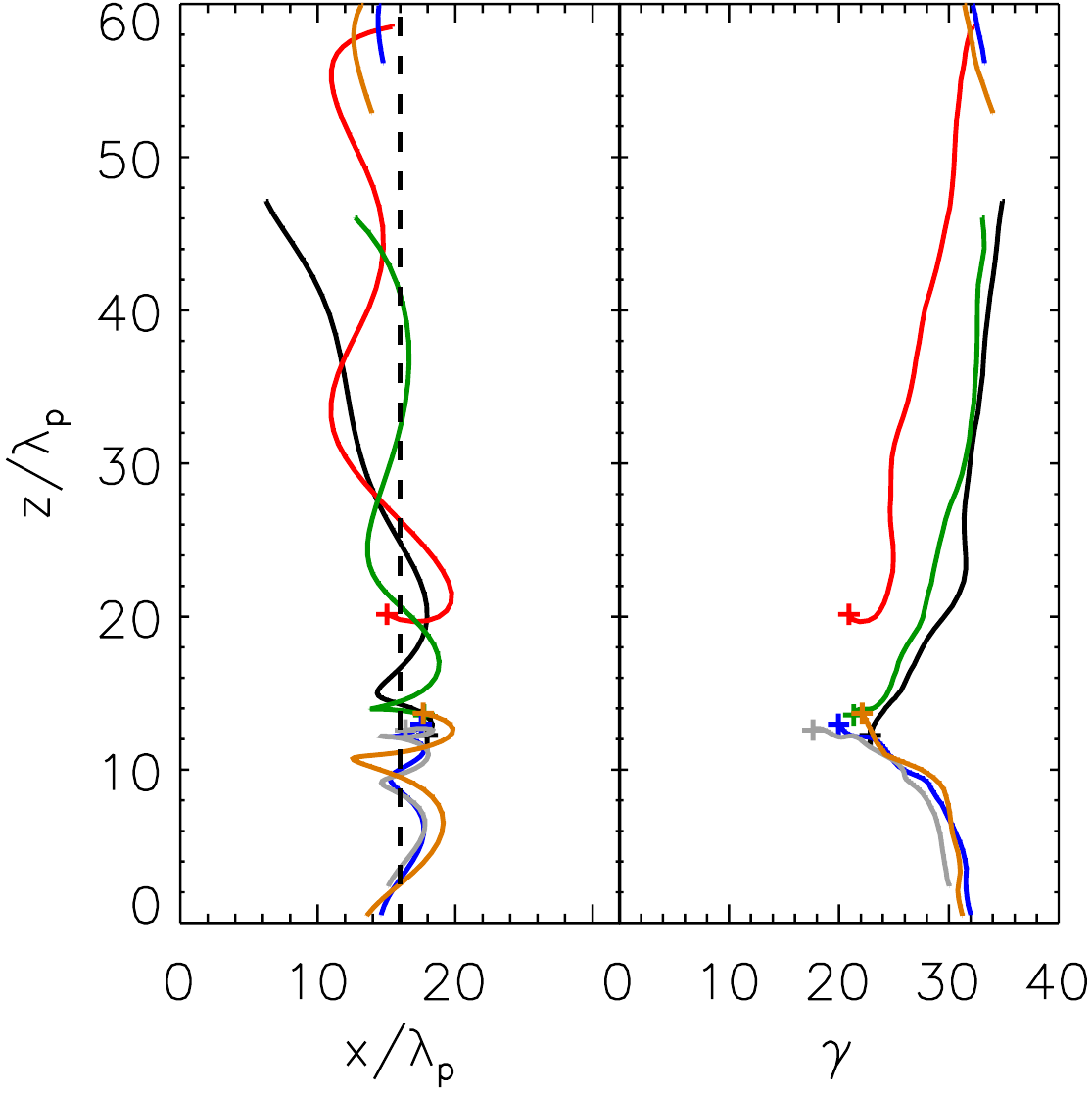}
\end{center}
\caption{The trajectories of six representative particles
that attain Lorentz factors $\gamma>30$ in the simulation {\tt
S2K025} (left panel), and the evolution of the Lorentz factor as a
function of coordinate $z$ (right panel).
The dashed line indicates the center of the current sheet. Note that
many of the particles start at $z\approx 12.5\,\lambda_{\rm p}$ and
close to the center of the current sheet; this is the location of the
large X-line on the lower left in Figure \ref{fig:acceleration}.  The crosses indicate the particle position at the beginning of the acceleration phase. \label{fig:trajplot} }
\end{figure}

In Figure \ref{fig:trajplot} we show the orbits of six representative
traced particles belong to the current sheet showing on the left of
Figure \ref{fig:acceleration}.  They all start in the current sheet within $\sim8\,
\lambda_{\rm p}$ from the X-line and have initial Lorentz factors in the range
$\gamma \sim 16-23$. While in the X-line region, the particles
oscillate across the current sheet on Speiser-like orbits.
The particles reach Lorentz factors of
$\gamma  \sim 32$ (this is how they were selected; see Figure \ref{fig:trajplot}, right panel) 
after drifting from the X-line region into the
neighboring islands.  
At the final Lorentz factor, the particles have
Larmor radii evaluated using the reversing magnetic field  $r_{\rm L} =
\gamma m_e c^2/(eB_0) \sim 17\,\lambda_{\rm p}$, larger than the width of the flux
ropes in the simulation. Because the magnetic field in  flux
ropes is only 
somewhat enhanced compared to the reversing field, the accelerated particles are not 
easily trapped within the flux ropes.

The electric field is nearly uniform across the
reconnection region, hence the electric force acting on a particle
trapped in a Speiser orbit is approximately constant in time provided
that the the inclination of the particle's velocity
relative to the direction of the electric force 
is relatively small.  Most of the traced particles happen to fulfill
this condition. Then,
the work done by the electric force on the particle is 
independent of the initial particle energy.  We have also checked
that, we expect for the Speiser orbit acceleration, it is the
$y$-component of the electric field that contributes the most to the
energy gain.  The average total energy gain per particle obtained by integrating
$\Upsilon$ over the volume of the simulation is $\Delta \gamma\approx 12$.

Because the rate of particle acceleration in the $y$-direction is
constant in time, the total energy gain is limited by the time the
particle spends in the X-line region \citep[see,
e.g.,][]{cerutti_extreme_2012}.   Particles displaced from the very
center of the X-line experience a Lorentz force due to the
reconnected magnetic field $B_x$ that deflects them away from the
X-line and toward the flux rope.  Once a particle has left the
X-line region and entered a flux rope, it no longer experiences
coherent acceleration.  
The lack of acceleration internal to flux ropes of the type proposed
by \citet{drake_electron_2006} was attributed in
previous works to the presence of the guide field
\citep{fu_process_2006, huang_mechanisms_2010}. Here, in addition the
presence of the guide field, we note that unlike the particles traced
in \citet{drake_electron_2006}, our energized particles
already have very large Larmor radii by the time they transition from
the X-line region into a flux rope and the magnetic geometry and
dynamics also seem markedly different.

\subsubsection{Angular Distribution of The Highest Energy Particles}

If the motion of high energy particles accelerated in a reconnection
region is anisotropic, then the synchrotron radiation emitted by these
particles carries angular dependence.  Radiation emitted by beams of accelerated particles
can come in and out of view of an observer; at any given time, the
observer detects only the radiation emitted by particles with momenta
making angles $\lesssim \gamma^{-1}$ from the line of sight.  Such
beaming could
explain the 
temporal variability often seen in astrophysical sources, such as GRBs
\citep[e.g.,][]{zhang_internal-collision-induced_2011,mckinney_reconnection_2012},
blazars \citep[e.g.,][]{nalewajko_radiative_2011}, that may be
powered by reconnection.  It could also influence the
characteristics of the gamma-ray emission from other candidate
reconnection-powered astrophysical sources, such as magnetars
\citep[][]{thompson_soft_1995,lyutikov_explosive_2003,parfrey_twisting_2012} and pulsar wind
nebulae \citep[e.g.,][]{lyubarsky_reconnection_2001,sironi_acceleration_2011,cerutti_extreme_2012}.  We investigate the angular distribution of
the momenta of the highest energy particles in our simulation, and
compare with the results of \citet{cerutti_beaming_2012}, who recently
reported a high degree of beaming in a two-dimensional PIC simulation
of pair plasma reconnection initialized with $\sigma\sim 40$.

\begin{figure*}
\begin{center}
\includegraphics[width = 0.85\textwidth]{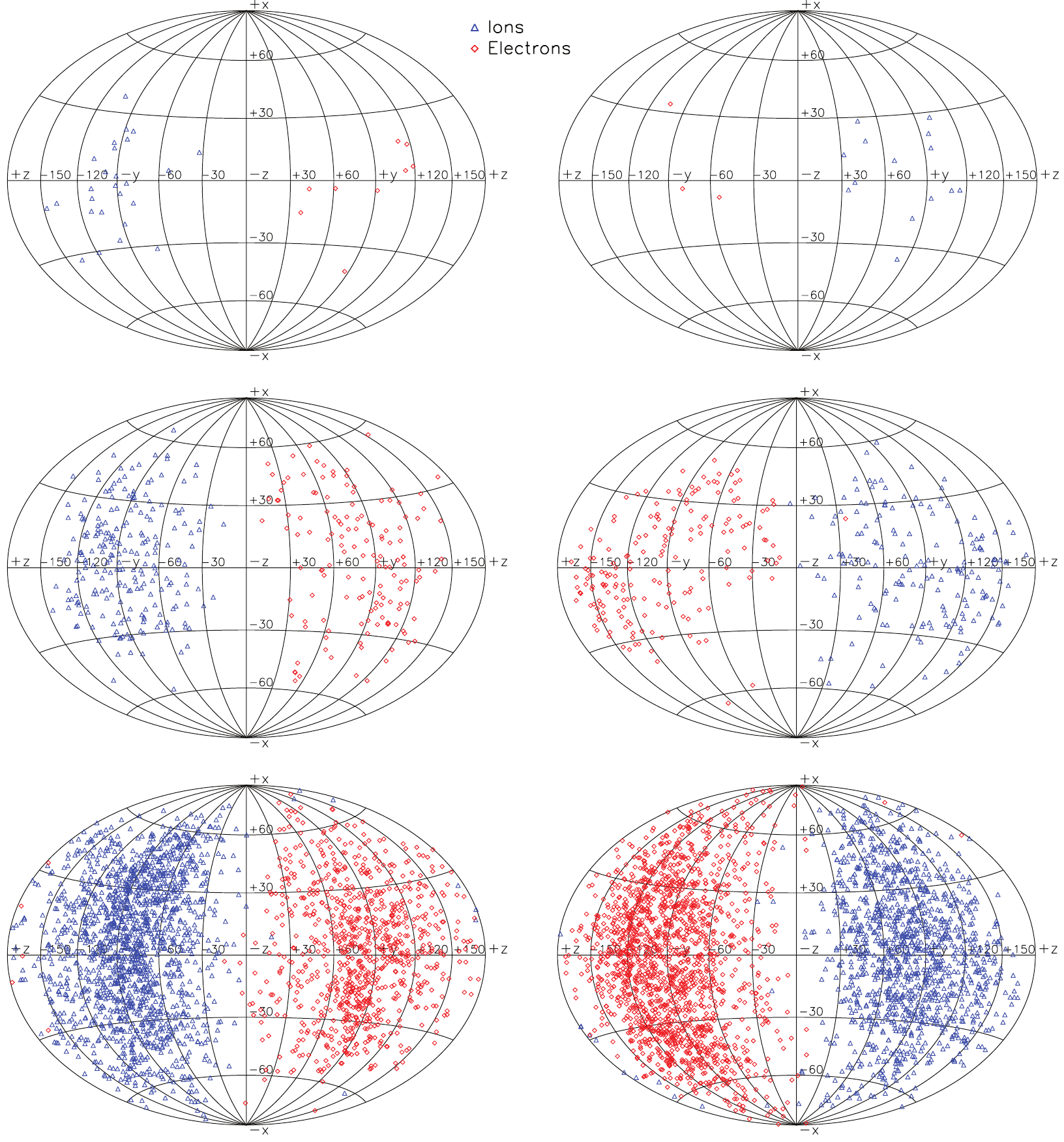}
\end{center}
\caption{Angular distribution of the momenta of particles with
  Lorentz factors $\gamma\geq 30$ at times $t=56\,\tau_{c0}$ (top panels), 
  $t=66\,\tau_{c0}$ (middle panels), and $t=84\,\tau_{c0}$ (bottom panels)  
  in simulation {\tt S2K025L} shown in the Aitoff projection. 
The panels of the left and right show the  particles
  near the current sheets at $x=L_x/4$ and $x=3L_x/4$,
  respectively.  The plotted particles were selected from a random sample containing $5\%$
  of the particles in the simulation. \label{fig:directions}}
\end{figure*}

Figure \ref{fig:directions}
shows the orientations of the momenta of all of the electrons and ions with
$\gamma \geq 30$ in simulation {\tt S2K025L}.  We select the larger
simulation for this analysis to include a larger range of three-dimensional effects to which
the particle anisotropy may be sensitive.  The median inclination of
electrons (ions) to the  $y$-axis (negative $y$-axis) is
$\sim30^\circ-40^\circ$, hence a half of particles in each charge
species occupies a fraction $\Delta\Omega/(4\pi) \sim 0.06-0.12$ of
the full solid angle. This moderate degree of beaming is a natural
consequence of X-line acceleration by the electric
field in the reconnection region, since the electric field is uniform
and accelerates the particles preferentially along the $y$-axis.

The degree of beaming, however, is much smaller than
in \citet{cerutti_beaming_2012}, where the highest energy particles,
with Lorentz factors $\gamma>40$,
were strongly beamed,  occupying a solid angle fraction as small as
$\Delta\Omega/(4\pi) \sim 0.01$.  The dependence of the beaming on the
parameters of the reconnection layer can crudely be understood as follows.
Assuming for simplicity that the plasma is ultrarelativistic and
highly magnetized, and that the energy of the accelerated particle
increases by a large factor, the components of its momentum following acceleration are approximately $p_y\sim \Delta \gamma m_e c$ and $p_x\sim p_z\sim\gamma_{\rm init} m_e c$, where $\gamma_{\rm init}$ is the initial Lorentz
    factor of an accelerated particle, and $\Delta \gamma$ is the
    Lorentz factor gain during acceleration, which increases the particle momentum in the $y$ direction. Therefore, the particle's degree of beaming is proportional to $\Omega/(4\pi)\propto (p_x^2 + p_z^2)/p_y^2 \propto(\Delta \gamma/\gamma_{\rm
      init})^{-2}$.   Using $\Delta \gamma
    \propto e E \Delta t/ (m_e c)$, where $E\sim r_{\rm rec} v_{\rm A}
    B_0 /c $ is the accelerating
    electric field (Equation \ref{recE}) and 
$\Delta t\sim \tau_{\rm merge} \sim \Delta z / v_{\rm A}$ is the
    duration of acceleration (Equation \ref{eq:tau_merge}), and expressing the magnetic field in
    terms of the magnetization parameter $B_0^2\propto \sigma n_0
    T_0\propto \sigma\left\langle \gamma\right\rangle\theta_0 (m_e c^2)^2/(e\lambda_{\rm p})^2$, where
    $\theta_0\equiv T_0/(m_e c^2)$ (Equations
    \ref{eq:sigma_uniform} and \ref{plasma}), we obtain
\begin{equation}
\label{eq:beaming_scaling}
\frac{\Omega}{4\pi} \propto \frac{1}{\sigma} \left(\frac{\gamma_{\rm  init}^2 }{
   \theta_0 \left\langle \gamma\right\rangle }\right) \left(r_{\rm rec}\frac{\Delta z}{\lambda_{\rm p}}\right)^{-2} .
\end{equation}
This result is accurate in the regime $\theta_0\gtrsim 1$ whereas Cerutti
et al.\ set up their simulation with $\theta_0=0.15$.  Ignoring this
concern, we
estimate that the factors in parentheses on the right hand side of
Equation (\ref{eq:beaming_scaling}) are similar in our simulation and
in Cerutti et al.   We therefore expect that the beaming solid angle
$\Omega/(4\pi)$ in their simulation should be smaller by the inverse
ratio of the 3s in the two simulations, which is
$(\sigma_{\rm Cerutti}/\sigma_{\tt S2K025L})^{-1} = 0.05$.  Thus, the
crude expectation is that the particles accelerated in Cerutti et al.\
should more beamed within a solid angle over an order of magnitude smaller
than in our simulation, consistent with the observed beaming ratio
between the two simulations.  In conclusion, this analysis suggests that high
magnetization and large physical size of the X-line region can both
give rise to a high degree of beaming.

\section{Discussion}
\label{sec:discussion}

\subsection{Tearing, Kink, and Oblique Modes}
\label{sec:mode_discussion}

A number of studies have examined the three-dimensional
evolution of a current sheet without a guide field; here, we focus on
results relevant to 
pair plasmas. \citet{zenitani_role_2008} found that the
kink mode dominated the tearing mode in antiparallel reconnection.  In
contrast, \citet{liu_particle_2011} found that the initial evolution
was dominated by the tearing mode, leading to an initial merging
phase similar to that in two dimensions.  After the completion of the
merging phase, however, they found that a secondary kink instability
set in and rendered the reconnection region turbulent and filled with
plasmoids covering a range of dimensions. The reduced influence of the
kink mode in the simulations of \citet{liu_particle_2011}
may be explained by initially relativistic drift velocities of
$0.82\, c$. Analytical calculations suggest that while the kink mode
dominates at low drift velocities, the tearing mode becomes
dominant for drift velocities exceeding $0.6 \, c$
\citep{zenitani_particle_2007}.  However, the growth rate of the
kink mode, unlike the tearing mode, depends
strongly on the initial structure of the current sheet
\citep{daughton_unstable_1999}.  

  In guide field reconnection, previous three dimensional studies have found that
  the kink mode is stabilized, but, under certain
  conditions, oblique modes dominate the evolution.  
 \citet{zenitani_role_2008} found that with a
  guide field present, the tearing mode and a sausage-like mode are combined into
  an oblique ``relativistic-drift-sausage-tearing''
  mode. 
\citet{daughton_role_2011} found that in electron-ion plasmas,
  oblique modes give rise to a network of interconnected flux ropes. 
   In both cases, the nonlinear development of the oblique
  modes led to turbulence in the current sheet. In nonrelativistic electron-ion
  simulations with a strong guide field, \citet{che_current_2011} 
found that at low temperatures, the current sheet
  filamented and that the turbulent transport
  played a key role in the diffusion of the field. At higher, albeit
  still nonrelativistic temperatures, their filamentation instability
  became weaker and the current sheet did not become turbulent.

One must be aware of the possibility
that unlike the tearing mode, the other instabilities identified in simulations starting with
Harris sheet equilibria are idiosyncrasies of the initial
configuration and have little to do with astrophysical magnetic
reconnection.  Our simulation, starting from a
non-Harris like configuration lacking a density enhancement in the initial
current sheet, should help discern artifacts of the initial configurations
from genuine properties of reconnection dynamics. 

We have found that in our simulations, in which we considered guide
fields at most equal to the reversing field, the tearing mode dominates the
evolution at all guide field strengths. To understand the difference between our simulations without guide field
and the studies that detected a dominant kink mode, we have carried out
two dimensional simulations with $\sigma=2$ and $\kappa=0$ in both the
$xy$ and $xz$ planes. We find that the kink mode grows faster
in a Harris current sheet, while the tearing mode grows faster
starting with our initial conditions. Our initial configuration seems
to enhance the growth rate of the tearing mode and inhibit the growth
of the kink mode.  The plasma inflow occurring during readjustment to
equilibrium may enhance growth of the tearing mode above the value
expected in pressure equilibrium.  Furthermore, our initial
configuration may be endowed with a somewhat stronger drift velocity
shear than the Harris sheet and this could stabilize the kink mode
\citep{volponi_shear-flow_2000} while having little effect on the
tearing mode \citep{roytershteyn_collisionless_2008}. 

In our simulations with guide field, we do not find any oblique modes similar to those observed by \citet{zenitani_role_2008} or \citet{daughton_role_2011}. Therefore, the only linear mode present is the tearing mode; the lack of phase coherence 
is the mechanism responsible for breaking translational symmetry. Because this 
mechanism is a general result of causal constraints, it is likely that 
this mechanism is present in Harris current sheets as well as in our simulations, 
but is masked by the more rapid effects of kink and oblique modes.

  \begin{deluxetable}{lccccc}
  \scriptsize
  \tablecolumns{6}
  \tablecaption{Tearing Instability Parameters\label{tab:tearing}}
\tablehead{\colhead{Run} & \colhead{$\beta_0$}  &\colhead{$f_{\rm c}$\,\tablenotemark{a}} & \colhead{$N_{\rm FR}$\,\tablenotemark{b}} & \colhead{$\,\tau_{c0}\omega_{\rm KTI,t} $\,\tablenotemark{c}}&\colhead{$\,\tau_{c0}\omega_{\rm KTI,s} $\,\tablenotemark{c}}}
\startdata 
{\tt S1K0}  &0.365&1.5&3&0.12&0.06\\
{\tt S1K025} &0.365&1.0&2&0.08&0.04 \\
{\tt S1K1} &0.365&1.0&2&0.08&0.03\\
{\tt S2K0}&0.344&2.3&4&0.16&0.14\\
{\tt S2K025}  &0.344&2.3&4&0.16&0.13\\
{\tt S2K025L}&0.344&2.3&9&0.16&0.11\\
{\tt S2K1} &0.344&2.3&2&0.09\tablenotemark{d}&0.07
\enddata
\tablenotetext{a}{$f_{\rm c} $ is the ratio of the original current sheet width to the current sheet width after the readjustment phase.}
\tablenotetext{b}{$N_{\rm FR} $ is the number of flux ropes per current
  sheet initially formed in the simulation.}
\tablenotetext{c}{The two values of $\,\tau_{c0} \omega_{\rm KTI} $ are
  the normalized theoretical (subscript ``t'') and simulated (subscript
  ``s'') growth rates of the tearing mode.}
\tablenotetext{d}{This growth rate is based on the observed mode wavelength, which is significantly longer than the fastest-growing wavelength.}
\end{deluxetable}

We proceed to calculate the theoretical and actual tearing mode growth
rates in each simulation. In Table \ref{tab:tearing} we show the two
rates which we label $\omega_{\rm KTI,t}$ and $\omega_{\rm KTI,s}$,
along with the initial particle drift velocity $\beta_0$, the initial
current sheet compression factor $f_{\rm c}$, and the number of flux
ropes per current sheet determined by visual inspection $N_{\rm FR}$.
 Because the fastest-growing tearing wavelength computing from the
 theoretical result in Equation (\ref{eq:tearing})  is $\sim10.8\,
 \lambda$ and the box size in all by simulation {\tt S2K025L} is
 $20\,\lambda_0$, 
the number of flux ropes per current sheet produced by the tearing
mode should be
$N_{\rm FR}=1.9 f_{\rm c}$; the number should be twice as
 large in {\tt S2K025L}.  The actual number of flux ropes formed in
 the simulation is agreement with this prediction, with the exception
 of the simulation {\tt S2K1} with the strongest guide field.  This run produced
 only 2 flux ropes per current sheet even though one expects 4.
 As discussed in \citet{daughton_role_2011}, the strong guide field
 may prefer oblique rather than pure tearing modes, but the former  
grow the fastest at long wavelengths $\sim 33\,
\lambda_{\rm p}$, about half the box size of the simulation, and the
finite box size interferes with ability of arbitrary oblique modes to grow.

  The values of $\omega_{\rm KTI,t}$ for the various runs are shown
  in Table \ref{tab:tearing}; it should be noted that these are upper
  limits to the possible tearing mode growth rates, because any measured
  growth rates are found from estimates in the nonlinear regime.  We
  also compute the simulated growth rates $\omega_{\rm KTI,s}$ of the
  tearing mode in our simulations by calculating the growth rate of the
  average reconnected magnetic field perturbation $B_x$ in the current sheet.  
Growth rates of the KTI calculated with the latter method should be
higher than those obtained by examining a single Fourier mode, as we
did in Section \ref{sec:mode_analysis}, because multiple Fourier modes
contribute to the amplitude of the perturbation. 
We find that the theoretical and simulated growth rates were relatively
    similar in all simulations, with theoretical rates being at most
    double the simulated rates. \citet{zenitani_particle_2007}
    identified a similar
    discrepancy between the theoretical and simulated growth rates. 
 The growth rates also show that decreased guide field $\kappa$, and
 especially higher magnetization $\sigma$, lead to higher tearing mode growth rates both in theoretical calculations and in the simulations. This reinforces the effects of the variation in $N_{\rm FR}$, which has a similar dependence on $\sigma$ and $\kappa$. 

We have shown that the tearing mode is responsible for producing magnetic
flux ropes in most of our simulations.  The tearing mode can also explain
many of the differences between the simulations.  The runs with higher
$\sigma$ and smaller $\kappa$ have larger values of $f_{\rm c}$,
which leads to higher tearing mode growth rates 
and the formation of a larger number of flux ropes $N_{\rm FR}$. The faster evolution of runs with higher $\sigma$ and smaller
$\kappa$ can be explained by the faster growth of the tearing mode in
such runs. Because runs with higher $\sigma$ and smaller $\kappa$ are
also associated with a larger $N_{\rm FR}$, more flux rope merging can take place
in such runs. Flux rope merging is strongly associated with energy
transfer and particle acceleration (as we show in the following
sections), so the larger energy transfer in runs with higher $\sigma$
and smaller $\kappa$ can be explained by the larger $N_{\rm FR}$ in
those runs. This suggests that a large portion of the variation
between runs is a result of the effect of $\sigma$ and $\kappa$ on the
tearing mode.

\subsection{Particle Energization}

Various particle energization channels have been identified in PIC
simulations of plasma reconnection \citep[see, e.g.,][and references
therein]{oka_electron_2010}. Three specific loci where
particle energization was detected include: inside or near the diffusion region
containing reconnection X-lines, in the magnetic islands (or
plasmoids, flux ropes) where the reconnected magnetic flux accumulates, and between an X-line and the edge of a flanking island, where the plasma flowing out of the X-line first encounters a strong magnetic field gradient. 

A number of studies point to the conclusion that 
significant energization
occurs near the primary (i.e., largest, or highest rank)
X-lines. There, the electric field, which is perpendicular to the
reversing field and is aligned with the current density vector in
the middle of the current sheet, can
accelerate particles oscillating within or across the current sheet
\citep[e.g.,][]{zenitani_generation_2001,zenitani_particle_2007,zenitani_role_2008,lyubarsky_particle_2008,uzdensky_reconnection-powered_2011,bessho_fast_2012,cerutti_extreme_2012,cerutti_beaming_2012}.
A variant of this mechanism involves the trapping of particles in
secondary magnetic islands appearing within the diffusion region of an
X-line \citep{oka_island_2010}; we will not discuss this as we do not observe secondary flux rope
formation in our simulations.

Particles can also be accelerated outside of the X-lines, such as 
in primary magnetic islands. If an island contracts in the course of
its relaxation to MHD equilibrium, the particles trapped inside it can be accelerated
by a Fermi-type process
\citep[e.g.,][]{drake_electron_2006,drake_magnetic_2010,kowal_magnetohydrodynamic_2011}.
Yet another location for particle acceleration is in the pileup region
between the X-line 
and a flanking island, which is where the reconnected magnetic
flux accumulates.
There, relativistic Speiser motion can combine with
curvature drift along the magnetic field gradient to create
significant acceleration \citep[e.g.,][]{hoshino_suprathermal_2001,jaroschek_fast_2004,zenitani_particle_2007,pritchett_energetic_2008,huang_mechanisms_2010,liu_particle_2011}.  
 
These mechanisms can operate in reconnection sites where the plasma
configuration, involving inflow into the diffusion region and outflow
toward the flanking islands, is relatively stationary, or inside a
single, autonomously evolving
magnetic island.  
There may be other mechanisms that operate in course of spatial
rearrangement and merging of magnetic islands.  Converging islands can
give rise to Fermi-type acceleration as particles bounce between them
\citep[e.g.,][]{oka_electron_2010,tanaka_dynamic_2011}. A dynamically
active reconnection region
containing many interacting islands can also allow Fermi-type stochastic
particle acceleration \citep{drake_magnetic_2010,hoshino_stochastic_2012}.  It is worth
noting that even if particle acceleration is not directly driven
by the merging, it is normally most active during island
merging episodes in the course of a reconnection event
\citep{jaroschek_fast_2004,pritchett_energetic_2008}. 

The presence of a guide field has a significant effect on particle
acceleration in reconnection \citep{zenitani_role_2008}. Without a guide field, the direct X-line
acceleration was typically found to be less efficient than the other
mechanisms discussed. However, with a guide field, the effectiveness
of acceleration in the pileup region \citep{huang_mechanisms_2010} and
inside magnetic islands \citep{fu_process_2006} was diminished, thus
leaving X-line acceleration of particles on Speiser orbits as the
dominant mechanism.  Our results are consistent with this conclusion.

To permit an interpretation of the nonthermal radiation spectra in
observed high-energy astrophysical sources in terms of synchrotron and
inverse-Compton radiation, a relatively hard power law particle energy
spectrum, which places a significant portion of the total energy in
high-energy particles, is required.  In most PIC simulations
including the present work, reconnection produced energized 
particle populations, but whether the populations represented genuine power laws
tails has remained a matter of interpretation.  Alternatively, an
energized population can be interpreted as a composite of one or more 
thermal sub-populations, each at a different temperature, for example.
Two dimensional simulations typically provide the dynamic range to
make a tentative distinction between a thermal or a power law spectrum
in an energized population. In most three-dimensional simulations,
however, such a determination is dubious. In what follows, we discuss
the characteristics of the particle energy spectra in the literature
and provide a comparison with our results.  Then, we briefly reflect
on the expected nature of particle energy spectra
produced by systems experiencing magnetic reconnection.

Most investigations involving PIC simulations of plasma reconnection
have interpreted a section of the high energy tail of the particle energy
spectrum as a power law $d
N/d\ln\gamma\propto \gamma^\alpha$. In the X-line region of two
dimensional simulations, spectra with power law indices as hard as $\alpha\approx-1$ have been reported
\citep{zenitani_generation_2001,zenitani_particle_2007,jaroschek_fast_2004,bessho_fast_2007,lyubarsky_particle_2008}. The
spectrum in the whole simulation box also contains a power law
component, but with a softer index of $\alpha\sim -2.5$. In their two-
and three-dimensional simulations of shock-induced reconnection,
\citet{sironi_acceleration_2011} find that as long as the region
containing a reversing magnetic field that
can undergo reconnection is reasonably large compared to the thickness of the
reconnection layer, the particle energy spectrum at high energies is a
power law with an index $\alpha=-1.5$ over a decade in energy.  Other
three dimensional investigations have identified  
relatively soft power law spectra with indices $\alpha\sim -3$ both in the X-line
region and in the entire simulation box
\citep{jaroschek_fast_2004,zenitani_role_2008}.  

Other investigations have interpreted particle energy spectra in terms of
multiple thermal and other exponentially 
truncated populations \citep[e.g.,][]{oka_electron_2010}. In their two-dimensional simulation
of reconnection in an initially nonrelativistic plasma at
temperature $T=0.15\,m_e c^2$,
\citet{cerutti_beaming_2012} find a new thermal population at
temperature $T\sim 4\,m_e c^2$. In
a three-dimensional simulation beginning with a relativistic
plasma at temperature $T=m_e c^2$, \citet{liu_particle_2011} find that
a thermal population with $T\approx2.3\,m_e c^2$ is produced.  
In a recent work
presenting two-dimensional simulations without a guide field,
\citet{bessho_fast_2012}
detect a spectrum of the form $dN/d\gamma\propto
\gamma^{-1/4}\exp(-a\gamma^{1/2})$, 
where $a$ is a constant of the order of unity.

Among the cited descriptions of particle populations energized by
reconnection, the spectra observed in our simulations bear resemblance
with those involving
multiple thermal sub-populations. For example, the spectrum at the end
of simulation {\tt S2K025L} can be modeled with three thermal
components at temperatures similar to those found in
\citet{liu_particle_2011}; specifically, the two energized
sub-populations have temperatures $T=2.1\,m_e c^2$ and $T=3.5\,m_e c^2$. 
Neither a power law nor the \citet{bessho_fast_2012} spectral form present
a good fit to the spectrum in this simulation.  It is important to
note that the continued evolution of the particle energy spectrum we observe
after the current sheet has become disordered, and the likely
additional evolution that would be taking place in am even larger
simulation, mean that the spectral form has not converged. For
example, it is possible that in addition to the two thermal components
that we have detected, additional such populations at still higher
temperatures would appear in larger simulations, and that the
combination of those would constitute a hard spectral tail,
e.g., a power law.

In PIC simulations, distinguishing between the various forms of energized particle spectra
is difficult due to dynamic range limitations. 
Another complication is how the spectrum of the non-energized, yet possibly
adiabatically heated background plasma is to be subtracted to isolate
the genuine nonthermal component.  Insight can separately be gained by
examining test particle trajectories in magnetic field geometries modeling
reconnection layers.  Particle ballistics in X-line as well as
magnetic island geometries
introduced at various levels of approximation has been investigated in
numerous studies \citep[e.g.,][]{zenitani_generation_2001,larrabee_lepton_2003,bessho_fast_2012,cerutti_extreme_2012}.  Often, the models entail transport terms describing particle
acceleration while confined in a single X-line region or an island, and
other terms describing particle escape and the termination of
acceleration.  Then, the terminal energy spectrum of the escaping
particles is obtained by taking the input energy spectrum and
deterministically transforming it by the transport terms. The crudest
such models have suggested that the terminal spectrum could be a power law,
e.g., of the form $dN/d\gamma\propto \gamma^\alpha$ with $\alpha = -
(2/\pi) B_{\rm rec}/E_{\rm RR} \sim -(2/\pi) c/v_{\rm A}$, where
$B_{\rm rec}$ is the reconnected magnetic field and $E_{\rm RR}$ is
the electric field in the reconnection region
discussed in Section \ref{sec:reconnection_rate}
\citep{zenitani_generation_2001}.  Models taking a more detailed
accounting of the transport of particle phase space coordinates and
the kinematics of escape,
however, can instead imply distinctly 
non-power-law, softer spectra \citep[e.g.,][]{bessho_fast_2012}.  It
is also important to note that the breaking of translational
invariance in the direction of the current, e.g., by three dimensional phase
decoherence (see Section \ref{sec:network}), may limit particle
acceleration and prevent the production of the same energized particle
tail produced in two dimensional geometries.

Power-law spectra are generically expected in acceleration
processes entailing stochasticity.  A prime example is the linear
diffusive shock acceleration (DSA) in which the random outcome of
particle scattering in the shock downstream determines whether the
particle will return to the shock upstream and be subjected to 
another acceleration cycle
\citep{bell_acceleration_1978,blandford_particle_1978}.  In contrast,
the acceleration in an idealized time-independent and two-dimensional X-line
region that lacks substructure in the form of secondary and embedded
islands,
should be deterministic.  Therefore the models of \citet{zenitani_generation_2001},
\citet{larrabee_lepton_2003}, and \citet{bessho_fast_2012}, 
can be thought of as producing a power-law-like spectrum only ``by coincidence.''
Realistic reconnection regions should plausibly allow particles to be accelerated in
stochastic fashion as they contain time dependence and structure on
multiple scales; they may also have
a mechanism for returning particles that have escaped an
acceleration site into another such site
\citep[see, e.g.,][]{drake_electron_2006,drake_magnetic_2010,kowal_magnetohydrodynamic_2011,hoshino_stochastic_2012}.
Since in this work we have found that the reconnection layer
transitions into a disordered network of interacting flux ropes, it will be
particularly interesting to investigate, in subsequent study, if this network
allows for stochastic acceleration of particles in intermittent, secondary reconnection sites that
may appear in the course of the flux rope network evolution.

\section{Conclusions}
\label{sec:conclusions}

In this paper, we carried out three dimensional PIC simulations of
magnetic reconnection in a relativistic pair plasma with varying guide
field strength.  Plasma magnetizations, expressed in terms of the
magnetic to kinetic pressure ratio, were of the order of unity.  The
initial conditions differed from the usual Harris sheet configuration
by not having a large density contrast between the center of the
current sheet and the background plasma.  We
investigated the growth of unstable kinetic modes in the current
sheet, as well as the nonlinear development of a three dimensional
flux rope network.  We also investigated the character and efficiency
of particle energization.  Our main results can be summarized as
follows:

The current sheets in all simulations develop significant magnetic
reconnection accompanied with conversion of magnetic to particle
kinetic energy.  With the aid of Fourier decomposition, we ascertained that the linear tearing mode is dominant
in the early evolution of the current
sheet.  We find that no significant growth occurs in the linear kink and oblique modes.
The nonlinear development of the tearing mode produces a chain of flux
ropes separated by primary X-lines.  The flux
ropes merge in hierarchical fashion whereby the
merging time scale is proportional to the flux rope separation. During
this phase magnetic reconnection takes place at the X-lines. We find that the dimensionless reconnection rates $\sim (0.05-0.08)$  and the maximum outflow speeds $\approx 0.4\, c \sim v_{\rm A}/2$ in our simulations are similar to those detected in other three dimensional simulations of reconnection in pair plasmas. We also find that spatial variation of an off-diagonal component of the pressure tensor is responsible for the breaking of flux freezing  at the X-lines, consistent with existing results.

While the hierarchical flux rope merging process 
initially appears similar to that found in two dimensional simulations, in fact it 
is three-dimensional from the outset.  This is because a 
lack of initial phase coherence in the linear tearing mode on scales
larger than those allowed by causality breaks translational invariance in
the direction of the initial current flow.  The flux ropes form a 
topologically interconnected, dynamically evolving network.
Dynamical interaction between neighboring flux ropes is provided by magnetic
tension forces. With time, the flux ropes break up into segments with more isotropic
orientations.  The strongly three-dimensional character of the
reconnection layer seems to suggest that global reconnection models invoking
quasi-two-dimensional plasmoid hierarchies 
\citep[e.g.,][]{shibata_plasmoid-induced_reconnection_2001,fermo_statistical_2010,uzdensky_fast_2010}
require revision to account for the inter-plasmoid magnetic linkage
and isotropization of plasmoid orientations. 

The larger flux ropes produced during flux rope merging contain
substructure down to plasma skin depth scales which is reflected in
embedded, twisted and braided current filaments and sheets.
Overall, this substructure is force-free and evolves relatively slowly.
However, isolated sites within the evolved flux rope network contain spatially
and temporally intermittent sites characterized by strong nonideal conditions ${\mathbf
  E}\cdot{\mathbf B}\neq 0$ where a change of magnetic
connectivity continues to take place even after flux rope merging has
saturated on length scales equal to the size of the computational
box.  This intermittency may produce the observed variability of
nonthermal emission in systems in which the emitting particles are
energized by magnetic reconnection.

During the early, ordered flux rope merging phase, particles are accelerated to high Lorentz factors by 
 the electric field in primary X-lines; the
trajectories of these particles are well described by Speiser
orbits. Particles continue to be energized in the later, disordered
phase we identify in our largest simulation, but we leave the analysis of energization in the disordered
regime to a subsequent investigation. 

Simulations with higher magnetization and lower guide field strength
exhibit greater and faster energy conversion and particle
energization.  The efficiency of particle energization measured in
terms of the energy in the accelerated particles per unit magnetic
energy dissipated in the simulation is an increasing function of the
guide field strength, which can be interpreted as resulting from a 
decreasing plasma compressibility with increasing guide field.
The final particle energy spectrum in the largest simulation is best fit by the inclusion of new
thermal components at temperatures $2.1\,m_e c^2$, and $3.5\,m_e c^2$,
in addition to the initial thermal component with temperature $m_e
\,c^2$.  We, however, acknowledge that a larger size or longer duration
simulation is likely to produce a still more pronounced energized
component, possibly even a population described with a power law spectrum.

Energetic positrons (electrons) with Lorentz factors $\gamma > 30$ are
moderately beamed in (opposite to) the direction of the initial current
flow with median inclinations of $\sim30^\circ-40^\circ$.
The degree of beaming is determined
by a particle's energy gain during acceleration.  We speculate that
more highly magnetized plasmas and reconnection sites with larger size
X-line regions should give rise to stronger beaming.

In this work, we have investigated a narrow range of magnetizations
with $\sigma\sim {\mathcal O} (1)$, but astrophysical reconnection
sites can also have high magnetizations $\sigma\gg 1$.  We can
speculate about the applicability of our results in the latter
limit. The linear tearing mode responsible for the initiation of
reconnection is insensitive to the degree of magnetization far from
the current sheet.  The phase decoherence that produces the initial
breakdown of translational invariance is determined by the tearing
mode growth time and should thus also persist at high
magnetizations. Therefore, we expect the qualitative structure of the
reconnection region at higher values of $\sigma$ to be similar to that
found in our simulations. 

The primary effect of high magnetization is that the Alfv\'en velocity
approaches the speed of light, which could give rise to
ultrarelativistic outflows from the X-line region. In such outflows the
inertial term of the generalized Ohm's law becomes important in the
breaking of flux freezing \citep{hesse_dissipation_2007}. This in turn
may increase the dimensionless reconnection rate $r_{\rm rec}$
relative to the value found in our simulations.  It remains to be seen
whether the associated reconnection process is more or less
intermittent.  An increased magnetization is likely to increase the
efficiency and the degree of beaming in particle energization.

We consider these results and the immediate questions they raise an
incremental step in the development of a multiscale view of collisionless
plasma self-organization during magnetic reconnection.  Further work
is clearly required to place our key finding, the evolution of the
simulated, periodic
reconnection layer into a disordered network of interacting magnetic flux
ropes, in the macroscopic context of a realistic reconnection site
characterized by outflow boundary conditions and altogether different
field line asymptotics at large distances from the X-line.   It will
be particularly interesting to see if the reconnected-flux-carrying
outflow from the macroscopic X-line will possess the disordered, interlinked
magnetic field topology we observe and investigate what will be the character of magnetic
fluctuations in the outflow. 

\acknowledgements
This work was initially inspired by Professor Pawan Kumar's
reflections on the physics of prompt emission in gamma-ray burst
sources and
we would like to thank Professor Kumar for many stimulating
conversations.
D.\ K.\ would also like to thank Professors Richard Fitzpatrick and Wendell Horton for useful discussions. 
A.\ S.\ is supported by NSF grant AST-0807381 and 
NASA grants NNX09AT95G and NNX10A039G.
The authors acknowledge the Texas Advanced Computing Center (TACC) at
The University of Texas at Austin for providing HPC resources that
have contributed to the research results reported within this paper.
M.\  M.\  and A.\  S.\ also gratefully acknowledge support and
hospitality from the Kavli Institute for Theoretical Physics during
their workshop Particle Acceleration in Astrophysical Plasmas
supported in part by the National Science Foundation under Grant No.\
PHY05-51164. We thank the Computational Information Systems Laboratory
at the National Center for Atmospheric Research for providing the
VAPOR analysis tool \citep{clyne_interactive_2007}.


\end{document}